\documentclass{aa}   
\usepackage[utf8]{inputenc}
\usepackage[T1]{fontenc}
\usepackage{graphicx}
\usepackage{fancyhdr}
\usepackage{nomencl}
\usepackage{setspace}
\usepackage{enumitem}
\usepackage{etoolbox}
\usepackage{amsmath}
\usepackage{amssymb}
\usepackage{physics}
\usepackage{mathrsfs}
\usepackage{natbib}
\usepackage[colorlinks]{hyperref}
\usepackage{algpseudocode}
\usepackage{algorithm}
\usepackage{booktabs}
\usepackage{subfigure}
\usepackage{dsfont}
\usepackage{xcolor}
\usepackage[font=small, skip=0pt, labelfont=bf]{caption}
\usepackage{stmaryrd}
\usepackage{diagbox}
\usepackage{mathtools}
\usepackage{mdframed}

\graphicspath{{./Images/}} 

\newcommand{\given}[1][]{\,#1\lvert\,}  

\addto\extrasenglish{
  \renewcommand{\sectionautorefname}{Section} %
  \renewcommand{\figureautorefname}{Figure} %

}

\let\orgautoref\autoref
\providecommand{\figRefShort}[1]{\def\figureautorefname{Fig.}\orgautoref{#1}\def\figureautorefname{Figure}}

\providecommand{\sectRefShort}[1]{\def\sectionautorefname{Sect.}\orgautoref{#1}\def\sectionautorefname{Section}}

\hypersetup{
    citecolor=[rgb]{0.04, 0.20, 0.58},  
    linkcolor=[rgb]{0.86, 0.078, 0.235},
    urlcolor=[rgb]{0,0,1}
}
\begin{document}

\title{T-ReX: a graph-based filament detection method}
\author{Tony Bonnaire\inst{1, 2}, Nabila Aghanim\inst{1}, Aurélien Decelle\inst{2}, Marian Douspis\inst{1}}
\authorrunning{T. Bonnaire, N. Aghanim, A. Decelle, M. Douspis}
\titlerunning{T-ReX}
\institute{
            Université Paris-Saclay, CNRS, Institut d'astrophysique spatiale, 91405, Orsay, France.
         \and
            Université Paris-Saclay, CNRS, Laboratoire de Recherche en Informatique, 91405, Orsay, France.
             }

\abstract{
    Numerical simulations and observations show that galaxies are not uniformly distributed in the universe but, rather, they are spread across a filamentary structure. In this large-scale pattern, highly dense regions are linked together by bridges and walls, all of them surrounded by vast, nearly-empty areas.
    While nodes of the network are widely studied in the literature, simulations indicate that half of the mass budget comes from a more diffuse part of the network, which is made up of filaments. In the context of recent and upcoming large galaxy surveys, it becomes essential that we identify and classify features of the Cosmic Web in an automatic way in order to study their physical properties and the impact of the cosmic environment on galaxies and their evolution.
    
    In this work, we propose a new approach for the automatic retrieval of the underlying filamentary structure from a 2D or 3D galaxy distribution using graph theory and the assumption that paths that link galaxies together with the minimum total length highlight the underlying distribution. To obtain a smoothed version of this topological prior, we embedded it in a Gaussian mixtures framework. In addition to a geometrical description of the pattern, a bootstrap-like estimate of these regularised minimum spanning trees allowed us to obtain a map characterising the frequency at which an area of the domain is crossed.
    Using the distribution of halos derived from numerical simulations, we show that the proposed method is able to recover the filamentary pattern in a 2D or 3D distribution of points with noise and outliers robustness with a few comprehensible parameters.
}


\maketitle


\section{Introduction} \label{Introduction}
    Large galaxy surveys like the Sloan Digital Sky Survey \citep[SDSS,][]{York2000} have confirmed the pattern drawn by  matter at very large scales, which was initially addressed in analytical works and the first N-body simulations \citep[e.g. ][]{Zeldovich1970, Shandarin1978} and which has also been exhibited in early observations \citep[see e.g.][]{Joeveer1978, Einasto1980}. In a pattern that is commonly referred to as the Cosmic Web \citep{Bond1996}, filaments act like cosmic highways, linking together large overdensities of matter and playing a key role in the dynamics of the universe. Since these early observations, the community has considerably enhanced the quality and the resolution of simulations with, for example, Millenium \citep{Springel2005}, Illustris \citep{Vogelsberger2014}, and Horizon-AGN \citep{Dubois2014}. These high-resolution simulations of dark matter (DM) evolution, which sometimes even include baryonic matter, have thus led to a more accurate spatial distribution of matter and allowed us to quantitatively characterise the different cosmic structures in terms of morphology, density, composition, etc. \citep[see e.g.][]{Colberg2007, AragonCalvo2010, Cautun2014, Gheller2016, Gheller2019}.
    Revealing the faint filamentary pattern of the Cosmic Web in data often relies on the view of galaxies as tracers of the dark matter distribution and allows for the study of the influence of the cosmic environment on the formation and evolution of those tracers \citep[e.g.][]{Alpaslan2014a, Alpaslan2014, Martinez2016, Kuutma2017, Malavasi2017, Laigle2018, Codis2018, Kraljic2019, Sarron2019, Malavasi2019}. It usually involves either stacking or individual inspection of objects after their detection. The observation of the filamentary pattern is currently performed using different observables: X-ray emissions \citep[see e.g.][]{Dietrich2012, Eckert2015, Nicastro2018}, weak lensing \citep[e.g.][]{Gouin2017, Epps2017}, or through the Sunyaev-Zel'dovich effect \citep[see e.g.][]{Bonjean2017, Tanimura2017, Degraaf2019, Tanimura2019}.
    
    To perform such statistical and physical analyses, it is essential to detect the filamentary pattern in an automatic way and this task is even more challenging when dealing with real observations. Visual inspection makes it possible to easily identify the underlying structure, especially in mock datasets, whether we are dealing with the filament-like or clustered parts of the pattern. Over the years, the key question quickly has shifted to how we can automatically extract that which is visually observed.
    In 1985, \citeauthor{Barrow1985} used, for the first time, a minimal spanning tree \citep[MST,][]{Boruvka1926} approach in a cosmological context to exhibit the underlying filamentary pattern from a 2D or 3D galaxy distribution, arguing that the usual statistical procedures, such as the two-point correlation function, are not sensitive to this specific feature.
    Since then, several methods have been developed to analyse and describe this gigantic network and yet, filaments still have not been attributed with a unique, well-posed definition. In an intuitive way, filaments correspond to bridges of matter between two dense regions of the space. On the basis of this simple idea, many algorithms with their own mathematical definitions have emerged. With no aim of being exhaustive \citep[see][for a detailed review]{Libeskind2017}, we give hereafter a list of such methods for classifying cosmic web elements.
    Some are using the previously mentioned minimum spanning tree, an object coming from graph theory. The resulting tree highlights a preferable path minimising the total distance to link galaxies together \citep[][]{Barrow1985, Alpaslan2014a}. After several processing stages of the graph proper to each method, filaments are extracted as branches of the tree.
    The study of the topological properties of the continuous density field through the Discrete Morse Theory led \cite{Spineweb} and \cite{DisperseTheory} to define filaments as the set of gradient lines linking maxima and saddle points.
    The seminal work of \cite{MMF} allowed \cite{Nexus} to build Nexus, an algorithm that performs a scale-space representation of the field in which filaments are defined locally through the relative strength between eigenvalues of the Hessian matrix of a smoothed continuous density obtained from the Delaunay Tessellation Field Estimator \citep{Schaap2000}.
    Another class of methods is based on a statistical representation of stochastic point processes to model the geometry of the filamentary structure. In particular, \cite{Stoica2007} presented their modeling of filaments as connected and aligned cylinders through the marked point-processes theory.
    \cite{Genovese2014} and \cite{Chen2015} proposed that cosmic filaments be identified as ridges in the distribution of galaxies using an automatic algorithm moving iteratively a set of points along the projected gradient.
    Some indirect methods aim to first recover the initial density field and then make it evolve forward in time using the Lagrangian perturbation theory. Indeed, \cite{Kitaura2012} and \cite{Jasche2013}, respectively, paved the way for \cite{Bos2014} and \cite{Leclercq2016} to develop such tools. We note that these methods are indirect reconstructions and are not specifically related to our issue of detecting cosmic web elements; although \cite{Leclercq2016} do use the inferred final density field in a game theory framework to classify structures in the reconstructed density field.
    
    This wide variety of approaches, all aimed at identifying filaments in a spatial distribution of matter tracers, reveals how this problem can be hard to handle and also how great an importance it holds for observational cosmology.
    Also, some of the above methods are designed on simulations and using dark matter particles to detect those features but if we want algorithms to be able to handle real datasets, we need it to work specifically with galaxies (or halos in simulation) as inputs.
    With this in mind, we developed an algorithm using a set of 2D or 3D galaxy positions to build a smooth representation given by a graph structure and standing in the ridges of the distribution. The presented method does not rely on any density estimation but directly on the set of observed data points. It does not assume any shape for filaments but, rather, a global weak prior on Cosmic Web connectivity and can be easily extended to any topological prior as long as it is given by a graph structure. Furthermore, it can be used as a denoised representation of the Cosmic Web for other applications than filament detection.
    
    
    In the first section, we present the datasets we use throughout this article to illustrate the steps and results of the proposed algorithm, called T-ReX (Tree-based ridge extractor). \autoref{Formalism} provides the required mathematical formalism used to build the procedure. \autoref{Detection} develops the method step by step and illustrate the obtained results on a simple dataset, while \autoref{Parameters} discuss the effect of each parameter on the resulting estimate of the underlying structure. Finally, \autoref{Results} presents and discuss outputs obtained on cosmological datasets, then comparing it with other existing methods, namely Bisous, DisPerSE, and Nexus.


\section{Data} \label{Data}
\begin{figure}
        \centering
        \begin{subfigure}
            \centering
            \includegraphics[width=1\linewidth]{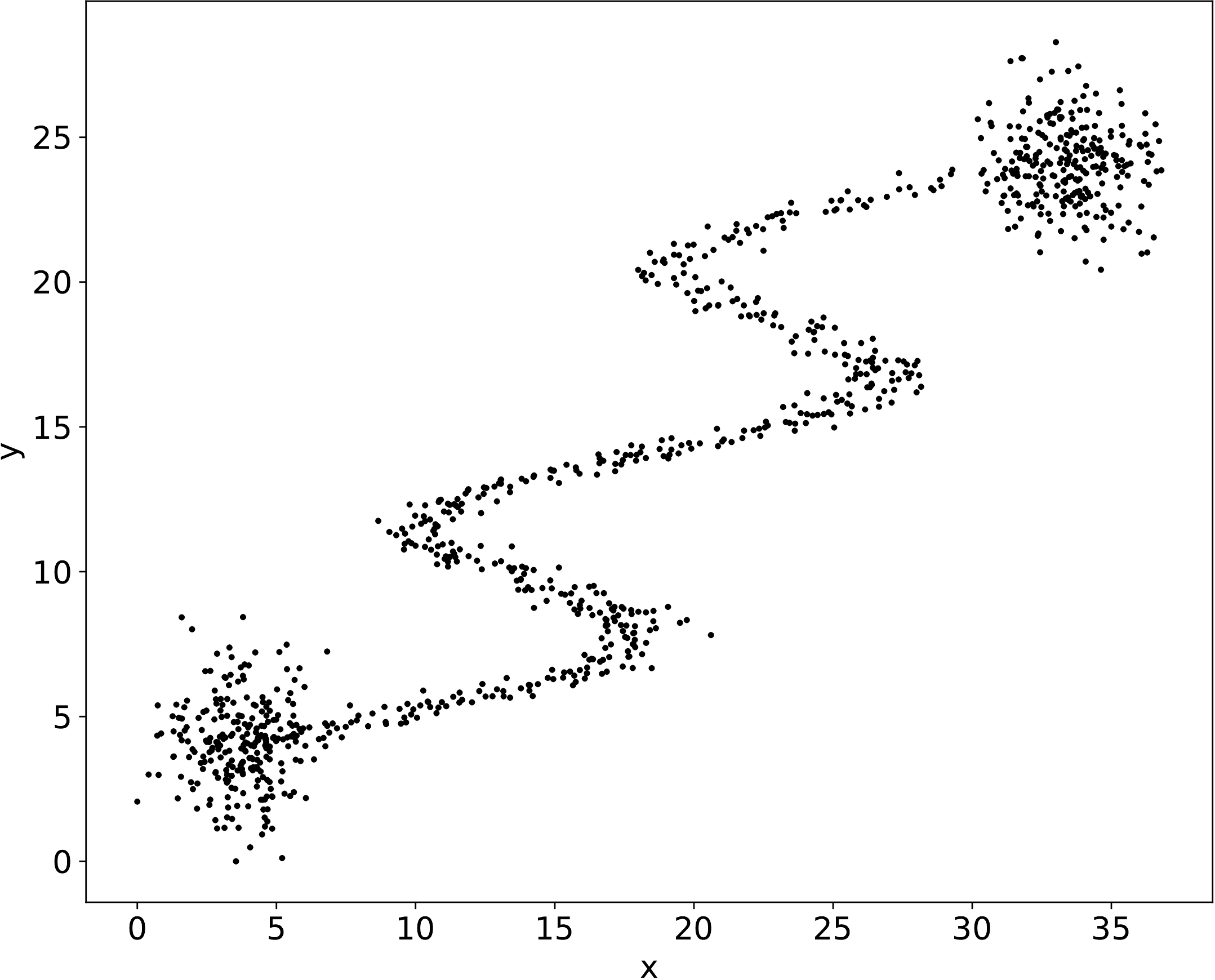}
        \end{subfigure}
        
        \caption{Toy model used to illustrate steps of the algorithm corresponding to a rotated sinewave with Gaussian random noise linking two Gaussian clusters.}
        \label{fig:ToyModel}
    \end{figure}

    In order to develop and test the main steps of the algorithm, we use a simple and non-cosmological dataset, hereafter called the toy dataset, shown in \figRefShort{fig:ToyModel}. It is constructed in a way so that it mimics a regularly curved structure, the filament, linking two clusters of points standing for overdense regions. The use of this toy dataset enables us to explore the impact of the parameters and test the reliability of the algorithm.
    
    \bigskip
    
\begin{figure}
        \centering
        \begin{subfigure}
            \centering
            \includegraphics[width=1\linewidth]{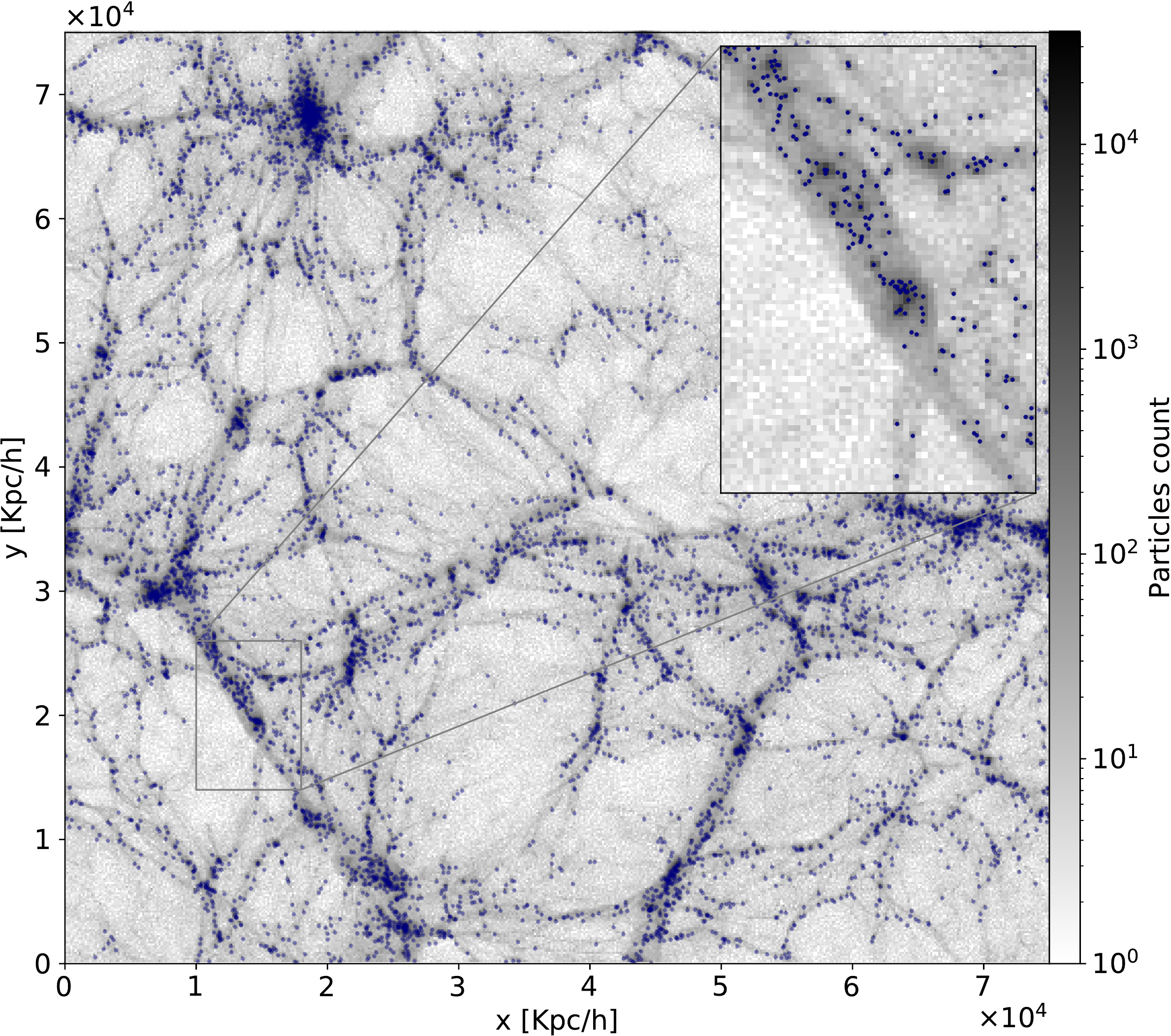}
        \end{subfigure}
        
        \caption{Projected 2D slice ($z_e=[0;5]\,$Mpc/h) of dark matter particles distribution obtained from Illustris-3 simulation at a redshift $z = 0$ together with 2D projection of Subfind subhalos in the same region (blue dots).}
      \label{fig:Illustris3}
    \end{figure}

    As a realistic cosmological dataset representing the Cosmic Web, we adopted the Illustris simulation outputs\footnote{\url{http://www.illustris-project.org/data/}} \citep{Vogelsberger2014}. It is a set of large-scale hydrodynamical simulations with different resolutions in which an initial set of particles (dark matter or baryonic gas) distributed over a $75$Mpc/h box is evolved forward in time from high redshift to $z=0$. From the resulting distribution at $z=0$, halos of dark matter are identified using a Friend-of-Friend algorithm \citep[FoF,][]{More2011}. To assess the application of the algorithm for cosmological cases and mimic its use for a galaxy survey, we consider structures inside halos, called subhalos, which have been identified with the Subfind algorithm \citep{Springel2008} and provided by the Illustris package, as has already been done in other recent studies \citep{Coutinho2016}. For convenience, we sometimes refer to these subhalos as 'galaxies'.
    \autoref{fig:Illustris3} shows a thin $5$Mpc/h slice of dark matter distribution obtained from the Illustris-3 simulation in which subhalos have been extracted. We can see how these 'galaxies' trace the underlying web drawn by the dark matter particles.
    
    In the following, each time we use a dataset built from the Illustris simulation, it always concerns the box at redshift $z=0$ and the Illustris-3 resolution obtained from $455^3$ dark matter particles with a mass resolution of $4.0\times 10^8 M_{\odot}$. When needed, we will explicitly specify the settings with which the subset of particles is obtained. Namely, we will specify the type of particles we are showing (subhalos or DM particles), the cut in the spatial distribution (over $x_e$, $y_e$ or $z_e$ spatial axes), and the cut in total mass $M$ over the considered particles in the spatial range.
    
    \bigskip
    
    Finally, to compare our results with other methods, we also apply T-ReX to FoF halos extracted from a $200$Mpc/h box of a Gadget-2 N-body simulation with $512^3$ particles \citep{Springel2005}.
    This particular simulation\footnote{\url{http://data.aip.de/tracingthecosmicweb/doi:10.17876/data/2017_1}} is the one used in \cite{Libeskind2017}, who proposed a unified comparison of the main existing procedures to classify elements of the cosmic web using either dark matter particles or dark matter halos as input.


\section{General formalism} \label{Formalism}

    Relying on the simple and only assumption that observed points (i.e. galaxies) are tracing the underlying Cosmic Web, the main idea of T-ReX is to model the filamentary structure as the set of ridges (or principal curves) in the input point cloud. To extract these ridges, we use the minimum spanning tree and extend its previous application in cosmology \citep{Barrow1985,Alpaslan2014} by building a smooth version of it standing 'in the middle' of the cloud. We note that this problem of finding curves passing through data points or detecting ridges in images is not proper to the cosmology field and has also been extensively studied in applied mathematics; it is currently of importance for medical applications, such as blood vessels segmentation \citep[see e.g.][for a recent review]{Moccia2018} or dimensionality reduction \citep{Qiu2017}.
    
    The basic idea behind this approach is that the true filamentary structure is a continuous manifold that can be described with a graph structure, while the observed galaxies represent a sparse and noisy sampling of that manifold. More precisely, in this paper is aimed at finding the best 1D representation of that manifold using a tree topology. This section introduces the required formalism to highlight how clustering methods as Gaussian Mixture Models (GMM), combined with graph theory, can be used to build such a representation starting from a general set of $N$ datapoints $X = \{x_i\}_{i=1}^N$ with $x_i\in \mathbb{R}^{d}$.
    
    \subsection{Elements from graph theory} \label{Graph_theory}
    Let $\mathcal{G}=(\mathcal{V}, \mathcal{E})$ be an undirected graph, with $\mathcal{V}$ as the collection of vertices, $\mathcal{E} = \{(i,j) \given (i,j) \in \mathcal{V}^2\}$ as the set of edges linking nodes together, and $\{w_{ij}\}_{(i,j)\in\mathcal{V}^2}$ as the set of edge weights, such that $\forall (i,j)\in\mathcal{V}, w_{ij} \geq 0$. In our case, we consider $w_{ij} = \lVert x_i - x_j \rVert^2_2$. Let us also define $d_i$ the degree of a node $i \in \mathcal{V}$ as the number of edges directly connected with it.
    
    We call minimum spanning tree\textit{} $\mathcal{M}$ the subgraph of $\mathcal{G}$ with $\abs{\mathcal{V}} - 1$ edges that is reaching all nodes of $\mathcal{V}$ with the minimum total weight. By construction, $\mathcal{M}$ has no loops and is unique if there are not two edges with the same weight in $\mathcal{G}$, which, in our case, does not seem likely to happen since it would imply galaxies with the exact same distance between them. Still, it would only create very local modifications of the tree structure that would be erased by future operations.
    In a tree-like structure, we can define three exclusive typologies for a node $i$ depending on its degree: extremity node ($d_i = 1$), junction node ($d_i = 2$), or bifurcation node ($d_i > 2$).
    
    Graphs can be represented by some computable quantities encoding the full graph information. A first representation is given by the adjacency
matrix\textit{} of $\mathcal{G}$, noted $\textbf{A}$, which is a symmetric $\abs{\mathcal{V}}\times \abs{\mathcal{V}}$ matrix encoding whether two vertices are linked or not. Elements $A_{ij}$ of this matrix take values as follows:
    \begin{equation}
        A_{ij} = 
        \left\{
        \begin{array}{ll}
            1 & \mbox{if } (i,j) \in \mathcal{E}, \\
            0 & \mbox{if } (i,j) \notin \mathcal{E}.
        \end{array}
        \right.
    \end{equation}
    This matrix encodes all the knowledge about the connectivity of vertices in the graph $\mathcal{G}$, and if we consider the matrix $\boldsymbol{W,}$ such that $W_{ij} = w_{ij}A_{ij}$, we end up with a matrix describing the full graph.
    
    Another useful representation of a graph is the Laplacian matrix\textit{}
 from which spectral decomposition gives fundamental information about the graph structure \citep{Lurie1999}. Let $\mathcal{G}$ be an undirected simple graph with an adjacency matrix $\boldsymbol{A}$ and $\boldsymbol{D}$ is a diagonal $\abs{\mathcal{V}}\times \abs{\mathcal{V}}$ matrix in which the element $D_{ii}$ corresponds to the degree of the node $i$. Then the Laplacian matrix of $\mathcal{G}$ is the symmetric, positive semi-definite $\abs{\mathcal{V}}\times \abs{\mathcal{V}}$ matrix defined as
    \begin{equation} \label{Laplacian}
        \boldsymbol{L} = \boldsymbol{D} - \boldsymbol{A}.
    \end{equation}
    
    \bigskip
    
\begin{figure}
        \centering
        \begin{subfigure}
            \centering
            \includegraphics[width=1\linewidth]{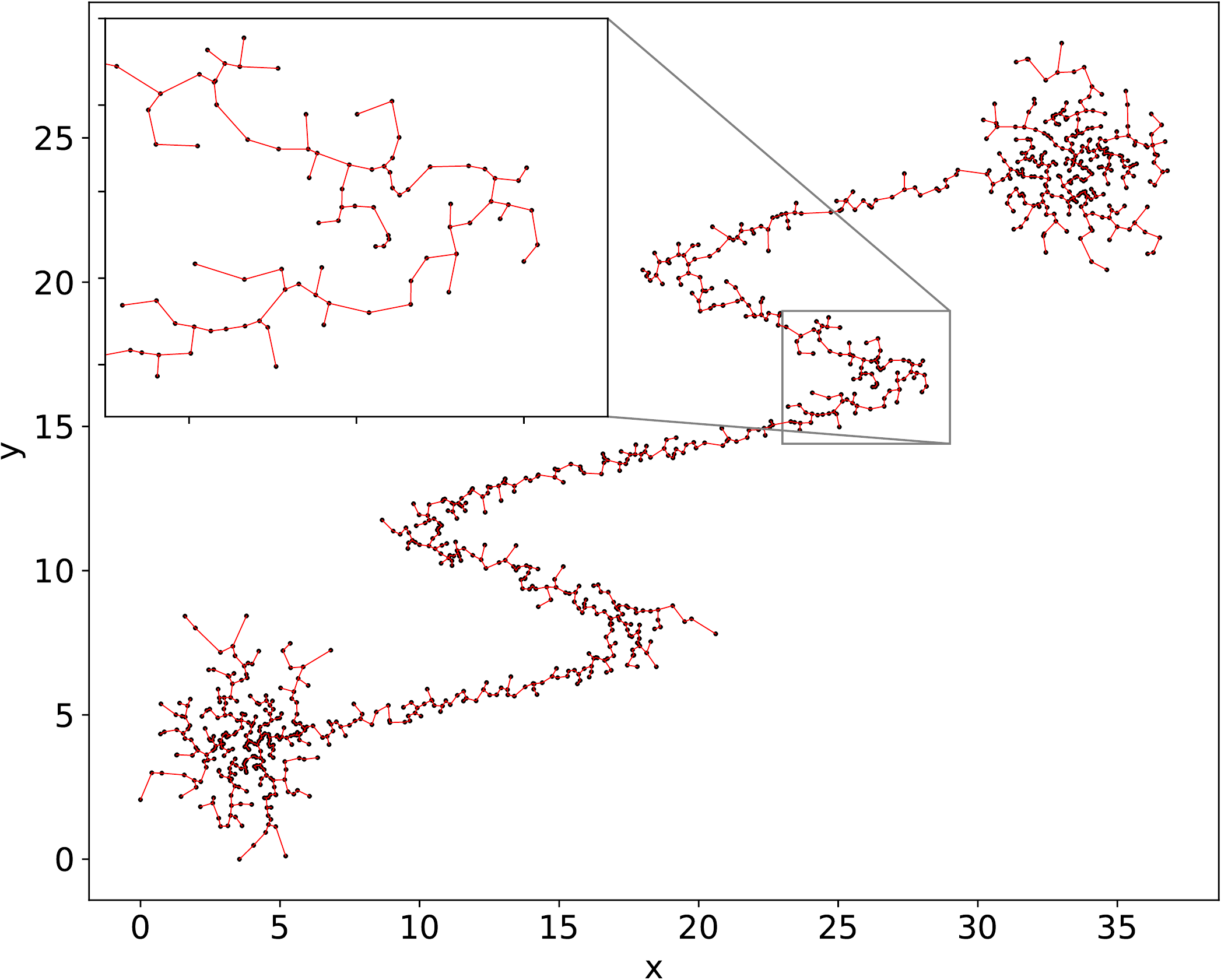}
        \end{subfigure}
        \caption{Minimum spanning tree computed over data points of the toy dataset. Black dots are data points and straight red lines are edges of the tree.}
        \label{fig:MST_toyModel}
    \end{figure}

    As the MST reaches all data points, the resulting graph is not smooth and, therefore, it does not properly reveal the local geometry of the underlying distribution (see \figRefShort{fig:MST_toyModel}). In order to recover the shape of the distribution, we span the set of data points with a given number of centroids that will coarse grain the density distribution. This task is achieved by using Gaussian Mixture Models. The key idea of T-ReX, thus, is to achieve a smooth representation of the d-dimensional dataset standing in its ridges by computing a set of centroids with an enforced topology given by a graph structure.
    
    \subsection{Expectation-maximisation for Gaussian Mixture Models}
    
    Gaussian Mixture Models (GMM) are part of parametric mixture models that can be used to map a cloud of points to a density distribution by using a restricted number $K$ of kernels to model the distribution. Starting with random parameters for Gaussian kernels, their positions and variances are adjusted iteratively to fit best the observed data. 
    GMM are also extensively used in unsupervised clustering approaches where the aim is to partition the datapoints into $K$ clusters by defining a probability that a given data point is part of the $k^\text{th}$ cluster. Using GMM, each cluster is represented by a Gaussian distribution and the clustering is reduced to an estimation problem of the Gaussian's parameters. Here we extend this second approach so that the clusters pave the observed set of datapoints in its ridges. 
    
    In practice, we define $K \leq N$ centroids $\{f_k\}_{k=1}^K$ with $f_k \in \mathbb{R}^d$ and assume that the dataset $X$ is drawn in an independant and identically distributed way from an unknown density that we model as a weighted linear combination of $K$ Gaussian clusters,
    \begin{equation} \label{eq:Mixture}
        p(x \given \theta) = \sum_{k=1}^K \pi_k \: \mathcal{N}(x \given f_k,\Sigma_k),
    \end{equation}
    where $\theta = \{\pi_1, \ldots, \pi_K, f_1, \ldots, f_K, \Sigma_1, \ldots, \Sigma_K \}$ is the set of model parameters, $\pi_k$ is the weight of the $k^\text{th}$ component, such that $\sum_{k=1}^K \pi_k = 1,$ and $\mathcal{N}(x \given f_k, \Sigma_k)$ is a multivariate normal distribution centered on $f_k$ with covariance $\Sigma_k$.
    
    This goal could also be achieved using the K-Means algorithm \citep{Macqueen1967} where we minimise the $L_2$ risk, 
    \begin{equation}
        R[f] = \frac{1}{N} \sum_{i=1}^N \min_{k=1\ldots K} \rVert x_i - f_k \lVert_2^2.
        \label{eq:Kmeans}
    \end{equation}
    This kind of similarity-based clustering of the data, however, generates a hard partition of the input domain, meaning that each point $x_i$ can only be member of one group $f_k$ and generally lacks of flexibility and robustness to noise and outliers. Mixture models can be used to face this difficulty by considering the conditional probability of a data point being part of a cluster given the assumed model.

    From the assumption that the data are drawn from such a density, all we have to do is to estimate the values for $\theta$ fitting best the observed data. This is generally achieved by maximising the log-likelihood function,
    \begin{equation}
        \mathcal{L}(\theta; X) =  \sum_{i=1}^N \log (\sum_{k=1}^K \pi_k \: \mathcal{N}(x_i \given f_k,\Sigma_k)),
    \end{equation}
    from which, in this case, it is impossible to get an analytic solution when maximising with respect to $\theta$.
    
    To bypass this difficulty, we use an Expectation-maximisation (EM) approach \citep{Dempster1977} by defining a set of latent variables, $Z = \{z_i\}_{i=1}^N$ , encoding the partition of the dataset: $z_i \in \llbracket1, K\rrbracket$ denotes which of the $K$ Gaussian components $x_i$ belongs to. The completed log-likelihood is then
    \begin{equation}
        \mathcal{L}(\theta; X, Z) =  \sum_{i=1}^N \log (\pi_{z_i} \, \mathcal{N}(x_i \given f_{z_i},\Sigma_{z_i})),
    \end{equation}
    which can be maximised using EM approach.
    
    As we introduced a new unknown quantity through $Z$, the central idea of the EM algorithm is to alternatively estimate $Z$ by the expectation over $p(z\given x)$ (E-step) and then update the parameters of the mixture $\theta$ by maximising the new likelihood on the basis of the current distribution for $Z$ (M-step). This procedure provides an algorithm that locally maximises the true likelihood.
    Mathematically, the procedure can be understood more generally as follows; for any probability distribution over the latent variables, $q(Z)$, it reads,
    \begin{equation}  \label{eq:Likelihood}
        \begin{split}
            \mathcal{L}(\theta; X) &= \sum_z q(z) \log (\frac{p(x, z \given \theta)}{q(z)}) \\ & \hspace{2cm} - \sum_z q(z) \log ( \frac{p(z \given x, \theta)}{q(z)})
            \\ 
            &= L(q, \theta) + D_{\text{KL}}(q \, \lvert\lvert \, p(z \given x, \theta)),
        \end{split}
    \end{equation}
    where $D_{\text{KL}}(q \, \lvert\lvert \, p) \geq 0$ is the Kullback-Leibler divergence \citep{KL1951}, implying that $L(q, \theta)$ is a lower bound for the log-likelihood.
    
    The idea behind EM formalism is to maximise the lower bound $L(q, \theta)$ instead of the log-likelihood directly. The E-step consists of fixing $\theta$ and maximising $L(q, \theta)$ with respect to $q$. By noting that $\mathcal{L}(\theta; X)$ does not depend on $q$, we simply need the divergence to be cancelled out in order to maximise the lower bound and, thus,
    \begin{equation}  \label{E_step}
        q(z) = \operatorname*{argmax}_{q(z)} L(q, \theta) = p(z \given x, \theta),
    \end{equation}
    which can be computed using Bayes' theorem.
    In the M-step, considering we are performing the $t^{\text{th}}$ iteration, we fix $q(z) = p(z \given x, \theta^{(t)})$ and update the optimal set of parameters, such that $\theta^{(t+1)} = \operatorname*{argmax}_\theta L(q, \theta)$.
    
    To summarise, EM is an iterative approach capable of identifying $K$ clusters from the data itself with guaranteed convergence. In a first step (E), a probabilistic (soft) assignment of each data point to mixture components is computed and in a second one (M) an estimation of mixtures' parameters is performed given the distribution for the latent variables. The main advantage over the K-means method is that GMM allow a soft partitioning of the input dataset through this $q(z)$ distribution.
    
    \subsection{Regularised GMM for ridge extraction} \label{RGMM}
    So far, we have simply addressed the Gaussian mixture clustering with an Expectation-maximisation approach and gained access to $K$ separated clusters, with their own means $\{f_k\}_{k=1}^K$ and covariances $\{\Sigma_k\}_{k=1}^K$ representing the data, but with no smoothness constraints or topology enforced. From the observation that the MST naturally traces ridges and the underlying connectivity of datapoints without any free parameters, we can enforce a tree topology to our centroids to obtain a representation that combines this idea of the MST and the local averaging naturally provided by GMM to impose smoothness. The question the full formalism tries to answer is what smooth minimal tree structure fits the set of observed data best.
    In general, if we want the centroids to have a given shape, we need to incorporate a prior distribution $p(\theta)$ within the previous equations. The presented framework is very close and inspired, in its form and spirit, to manifold learning methods for dimensionality reduction \citep[see e.g.][]{Roweis2000, Elmap2005} and, in particular, the principal curves \citep{Hastie1989} field, which has already studied the application of mixture models to curve extraction from point distribution \citep{Tibshirani1992, Bishop1998}.
    
    With such a prior, we no longer aim to directly maximise the likelihood but the posterior $\log p(\theta \given x) \propto \mathcal{L}(\theta; X) + \log p(\theta)$. In this context, previous equations and results from EM algorithm remain unchanged for the E-step, the maximisation over $q$ being independant on $p(\theta)$. In the case of the M-step, the update is computed so that $\displaystyle \theta^{(t+1)} = \operatorname*{argmax}_\theta L(q, \theta) + \log p(\theta)$.
    
    The log-prior can be considered as a regularization term on the log-likelihood and keeping in mind its role helps us choosing it correctly. In particular, we want to give centroids a smoothness constraint and to enforce a topology through a given graph structure $\mathcal{G}$. Hence, we use a Gaussian form for the prior with a variance $\nu^2$ thus acting on the $L_2$ norm $\lVert \boldsymbol{F} \rVert^2_\mathcal{G}$ to constrain the smoothness of centroids directly on the graph domain, as is usually done in statistics \citep{Smola2001} and which is inspired by previous studies on elastic topology regularization \citep{Durbin1987, Yuille1990} and manifold learning \citep{Elmap2005}:
    \begin{equation}
        \begin{split}
            \log p(\theta) &= - \frac{1}{2} \sum_{i=1}^K \sum_{j=1}^K b_{ij} \frac{\lVert f_i - f_j \rVert^2_2}{\nu^2} + \text{cte}, \\
            &= - \frac{1}{\nu^2} \Tr{\boldsymbol{F} \boldsymbol{L}\boldsymbol{F}^T} + \text{cte}. 
    \end{split}
    \end{equation}
    where $\boldsymbol{F} \in \mathbb{R}^{d\times K}$ such that column $k$ of $\boldsymbol{F}$ contains $f_k$ and $L$ is the Laplacian matrix as defined in Eq. \eqref{Laplacian}.
    
    \bigskip 
    In the context of this paper and its application, we simplify this formalism by considering equidistributed Gaussian mixtures ($\forall k \in \llbracket1, K\rrbracket, \pi_k = 1/K$) with identical and isotropic covariances $\sigma^2 \boldsymbol{I}_{d}$, where $\boldsymbol{I}_d$ denotes the $d\times d$ identity matrix.
    This reduces the problem with regard to the estimate of $\theta = \{f_k\}_{k=1}^K$ during the M-step.
    By noting $p_{ik} = p(z_i = k \given x_i, \theta_k),$ the probability of a given data point $x_i$ being well represented by the cluster $k$, we find
    \begin{equation}
        \begin{split}
            \theta^t &= \operatorname*{argmax}_\theta - \sum_{i=1}^N \sum_{k=1}^K p_{ik} \frac{\lVert x_i - f_k \rVert_2^2}{\sigma^2} \\
            & \hspace{3cm} - \sum_{i=1}^K \sum_{j=1}^K b_{ij} \frac{\lVert f_i - f_j \rVert^2_2}{\nu^2}.
        \end{split}
    \end{equation}
    The first term of this optimisation problem corresponds to a soft K-means clustering \citep{Bezdek1981} while the right-hand side is an elastic regularization term constraining the topology of centroids.
    Under the previous simplifications and in pursuit of a specific topology given by the minimum spanning tree, the presented formalism is equivalent to the work of \cite{Mao2015}.
    
    Again, to simplify the notation and to link the two variances $\sigma^2$ and $\nu^2$, we can introduce the parameter $\displaystyle \lambda = \frac{\sigma^2}{\nu^2}$ as the relative strength of the two kernels. The final problem of the M-step can hence be written as
    \begin{equation}    \label{RGMM_Mstep}
        \begin{split}
            \theta^t &= \operatorname*{argmin}_\theta \sum_{i=1}^N \sum_{k=1}^K p_{ik} \lVert x_i - f_k \rVert_2^2 \\
            & \hspace{3cm} + \lambda \sum_{i=1}^K \sum_{j=1}^K b_{ij} \lVert f_i - f_j \rVert^2_2.
        \end{split}
    \end{equation}
    The first term of this equation tries to minimise the error when datapoints are approximated by centroids while the second term acts like an elastic constraint on centroids when they are linked together in the considered graph. $\lambda$ can be seen as a regularization parameter acting like a soft constraint on the total length of the graph and, thus, as a trade-off parameter between the data fidelity term and the penalty term constraining the smoothness and simplicity of the graph representation.


\section{T-ReX: Tree-based Ridge eXtractor} \label{Detection}

    Given a set of $N$ observed data points $X = \{x_i\}_{i=1}^N$, each living in a d-dimensional euclidean space $\mathbb{R}^d$, the first step of T-ReX is to build a graph with a tree structure. This is achieved by computing the MST over $X,$ resulting in a unique preferable path to link points together (see \figRefShort{fig:MST_toyModel}).
    This tree then goes through several processes to obtain a version that is robust to noise and outliers and to gain some smoothness properties.
    
    \subsection{Pruning of the tree} \label{Pruning}
    
    Considering that we obtained a graph with a tree structure, we adopt a simple denoising operation by cutting all the nodes standing in branches of the tree at a level $l$. In practice, branches are defined as the set of connected nodes linking an extremity node to a bifurcation node (defined in \sectRefShort{Graph_theory}). Strictly speaking, we iteratively remove all nodes of degree one in the graph structure. By doing so, we remove the most spurious part of the structure corresponding to nodes that are more likely to be found in physically irrelevant regions for the underlying pattern (i.e. underdense regions). This approach is iterative, meaning that nodes which are initially bifurcations can become junctions or extremities (or even be removed if there are only branches with path length strictly lower than $l$ connected to it). To give a representative image of this procedure, it acts like iterative peeling of an onion, attributing to each node a depth in terms of layers to peel before we reach it and starting from extremities \citep{Hebert-Dufresne2016}.
    This method is very close to the first step introduced by \cite{Barrow1985}, where all branches with a path length inferior to $l$ are removed (meaning that there are less than $l$ nodes in the branch) except that our approach also cuts extremities of longer branches.
    
    Previous MST methods usually perform, in addition to this pruning, a removal of all edges above a given physical length. In our case, this operation is not only aimed at avoiding the introduction of a new parameter that is not easy to tune, but it is also based on  our argument that all connections, even 'long' ones, can provide information about the underlying structure. Of course,  as a result, if two unconnected parts of a network are given as an input to the presented method, they will end up connected.
    
\begin{figure}
        \centering
        \begin{subfigure}
            \centering
            \includegraphics[width=1\linewidth]{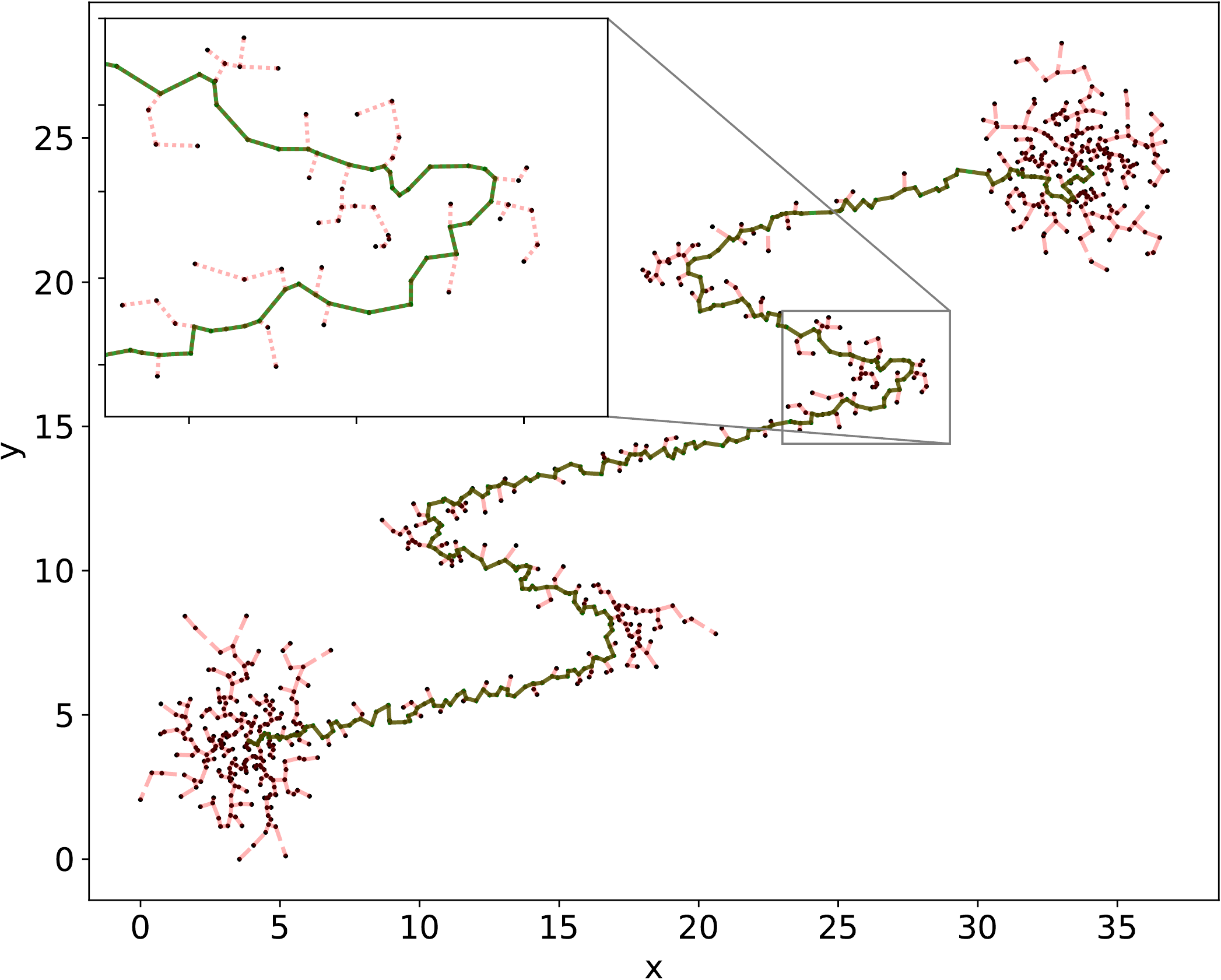}
        \end{subfigure}
        \caption{Pruned version of the minimum spanning tree displayed on \figRefShort{fig:MST_toyModel} at level $l=28$. Black dots are data points, dashed shaded red lines are edges of the MST and green solid lines are the remaining edges after pruning.}
        \label{fig:Pruned_toyModel}
    \end{figure}

    \autoref{fig:Pruned_toyModel} shows the pruned MST obtained with a given cut-off level on the toy dataset. We can clearly observe that removing extremity nodes iteratively acts like a denoising operation, deleting small branches and irrelevant ones while preserving the core of the pattern. The choice of the pruning level is essential for a single realization of a tree, especially when dealing with noisy data. \autoref{Parameters} analyses the impact of this parameter on the resulting tree.
    
    \subsection{The regularised minimum spanning tree} \label{RMST}
    As discussed in \sectRefShort{Formalism}, the MST does not exhibit a smooth behaviour. To enforce this constraint in our representation, we solve expectation-maximisation Equations \eqref{E_step} and \eqref{RGMM_Mstep} following the work and notations of \cite{Mao2016} by applying \autoref{Algo:PPT}. It is worth noting that the inverse of $2\lambda\boldsymbol{L} + \boldsymbol{\Lambda}$ always exists since $\boldsymbol{L}$ is a positive semi-definite matrix and $\boldsymbol{\Lambda}$ is a positive diagonal matrix. The convergence is guaranteed by the EM approach and characterised by a slow displacement of the projected points $\lVert \boldsymbol{F}_t - \boldsymbol{F}_{t-1} \rVert^2_2 \leq \epsilon$ where $t$ denotes the iteration index.
    
    The computational complexity of \autoref{Algo:PPT} can be divided into three components: i) The computation of the MST over the centroids, ii) The computation of the assignment matrix $\boldsymbol{P}$ to solve the E-step, and iii) The matrix inversion to update centroids positions during the M-step. As already pointed out in \cite{Mao2015}, the total complexity is $\mathcal{O}(K^3 + DNK + K^2D)$.
    
    \begin{algorithm}
      \caption{Regularised minimum spanning tree}
      \label{Algo:PPT}
      \begin{algorithmic}
        \Statex \textbf{Input:} Data: $\boldsymbol{X} \in \mathbb{R}^{d\times N}$, parameters: $\lambda$ and $\sigma$
        \Statex \textbf{Output:} $\boldsymbol{F} \in \mathbb{R}^{d\times K}$, the set of centroids and $\mathcal{B}$, the associated adjacency matrix
        \\
        \State Initialise $\boldsymbol{F} = \boldsymbol{X}$ or with K-Means clustering
        \\
        \While{convergence}
            \State Compute the minimum spanning tree $\mathcal{B}$ from $\boldsymbol{F}$
            \State Compute the Laplacian matrix $\boldsymbol{L}$ of $\mathcal{B}$ via equation \eqref{Laplacian}
            
            \State \textbf{E-step:}
            \State \quad Compute the assignment matrix $\boldsymbol{P}$ where $(i,k)$ entry is 
            \begin{equation}
            p_{ik} = \frac{\exp(-\frac{1}{2\sigma^2} \lVert x_i - f_k \rVert^2_2)}{\sum_{j=1}^K\exp(-\frac{1}{2\sigma^2} \lVert x_i - f_{j} \rVert^2_2)}
            \end{equation}
            
            \\ \State \textbf{M-step:}
            \State \quad Compute $\boldsymbol{\Lambda}$, a diagonal $K\times K$ matrix such that $\Lambda_{kk} = \displaystyle \sum_{i=1}^N p_{ik}$
            \State \quad Solve equation \eqref{RGMM_Mstep} to update the position of centroids\footnotemark, $\boldsymbol{F} = \boldsymbol{XP}(2\lambda \boldsymbol{L} + \boldsymbol{\Lambda})^{-1}$ 
            
        \EndWhile
      \end{algorithmic}
    \end{algorithm}
    \footnotetext{In term of these matrices, optimization problem \eqref{RGMM_Mstep} can be written  $\displaystyle \operatorname*{argmin}_{\boldsymbol{F}} \Tr{\boldsymbol{F} \boldsymbol{\Lambda} \boldsymbol{F}^T - 2 \boldsymbol{X} \boldsymbol{P} \boldsymbol{F}^T + 2\lambda \boldsymbol{F} \boldsymbol{L} \boldsymbol{F}^T}$}
    
\begin{figure}[t]
        \centering
        \begin{subfigure}
            \centering
            \includegraphics[width=1\linewidth]{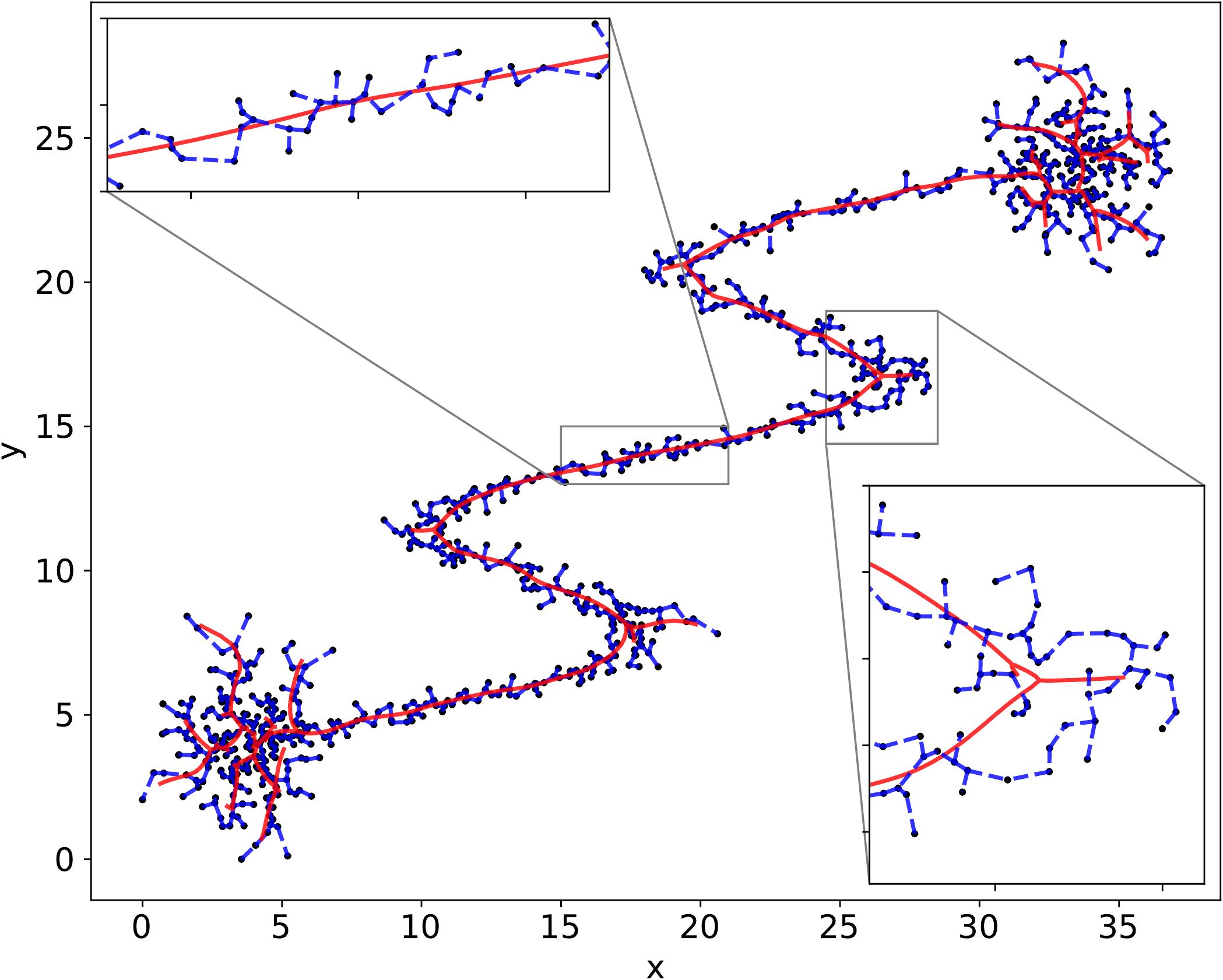}
        \end{subfigure}
        \caption{Regularised minimum spanning tree computed over data points of the toy dataset. Black dots are data points, red solid lines are edges of the regularised tree and dashed blue line is the original MST. Result obtained from \autoref{Algo:PPT} with $\lambda=1$ and $\sigma^2=0.67$ (explained in \sectRefShort{Parameters}).}
        \label{fig:RMST_toyModel}
    \end{figure}

    \autoref{fig:RMST_toyModel} shows the difference between the MST directly built on data points and its regularised version obtained from \autoref{Algo:PPT}. The regularised minimum spanning tree (RMST) has smooth extensions (visible in the zooms of \figRefShort{fig:RMST_toyModel}) while preserving the global shape of the tree-like structure. In the inflexion regions of the filament, we observe that the tree is a creating bifurcations. This is due to the chosen MST topology for the centroids. In this precise case, with a single filament, the best topology for centroids would be a straight line described by an adjacency matrix such that $A_{ij} = 2\delta_{i,i} - \delta_{i,j+1} - \delta_{i,j-1}$, where $\delta_{i,j}$ denotes the Kronecker delta function. It should be noted that such a topology could be handled by the formalism presented in \sectRefShort{RGMM}.
    
    \subsection{The probability map} \label{FrequencyMap}
    
    As previously mentioned in \sectRefShort{Formalism}, a graph with a tree structure has no loops and, hence, it cannot represent holes but only, rather, connected components in the Cosmic Web topology. In addition to that, the MST highlights one particular path linking data points together but does not provide any idea of uncertainty or reliability of this latter.
    Both of these issues can be overcome by introducing a robust representation that takes into account the eventual variations in the input distribution. To do so, we build $B$ different samples $\{X_b\}_{b=1}^B$ from the initial one $X$ and compute the regularised MST for each of them in a similar fashion as in bootstrap approaches.
    The entire procedure is described by \autoref{Algo:Frequency_map}.
    
    From the $B$ realisations of RMST, one can build a map $\boldsymbol{I}$ characterising the probability, in a frequentist meaning, of a position $x$ to be crossed by a realization of a tree:
    \begin{equation} \label{eq:FreqMap}
        \boldsymbol{I}(x) = \frac{1}{B} \sum_{b=1}^{B} 1_{\boldsymbol{H}_b(x)=1},
    \end{equation}
    where $1_{A}$ is the indicator function and $\boldsymbol{H}_b$ is the binary histogram obtained from the projected points $\boldsymbol{F}_b$. The random nature of $\boldsymbol{I}$ thus comes from the uniformly at random resampling of $X$ and not from \autoref{Algo:PPT} that is a deterministic optimization step.
    
    \begin{algorithm}
      \caption{Bootstrap RMST}
      \label{Algo:Frequency_map}
      \begin{algorithmic}
        \Statex \textbf{Input:} Data $\boldsymbol{X}$, parameters $\lambda, \sigma, l, B, N_B$
        \Statex \textbf{Output:} $\boldsymbol{S}$, the set of points describing the skeleton
        \State Generate $B$ bootstrap samples $\{\boldsymbol{X}^b\}_{b=1}^B$ of size $N_B$
        \For{each $\boldsymbol{X}^b$}
        \State Compute the MST $\mathcal{B}_b$ of $\boldsymbol{X}^b$
        \State Prune $\mathcal{B}_b$ at level $l$
        \State Keep the remaining vertices in $\mathcal{B}_b$, noted $\boldsymbol{Y}^b$
        \State Apply Algorithm \ref{Algo:PPT} on $\boldsymbol{Y}^b$ with parameters $\lambda$ and $\sigma$ to obtain the regularised MST $\mathcal{B}_b^R$ and optimal $\boldsymbol{F}_b$
        \EndFor
        \State  $\boldsymbol{S} = \{\boldsymbol{F}_b\}_{b=1}^B$
      \end{algorithmic}
    \end{algorithm}
    
\begin{figure}[b]
        \centering
        \begin{subfigure}
            \centering
            \includegraphics[width=1\linewidth]{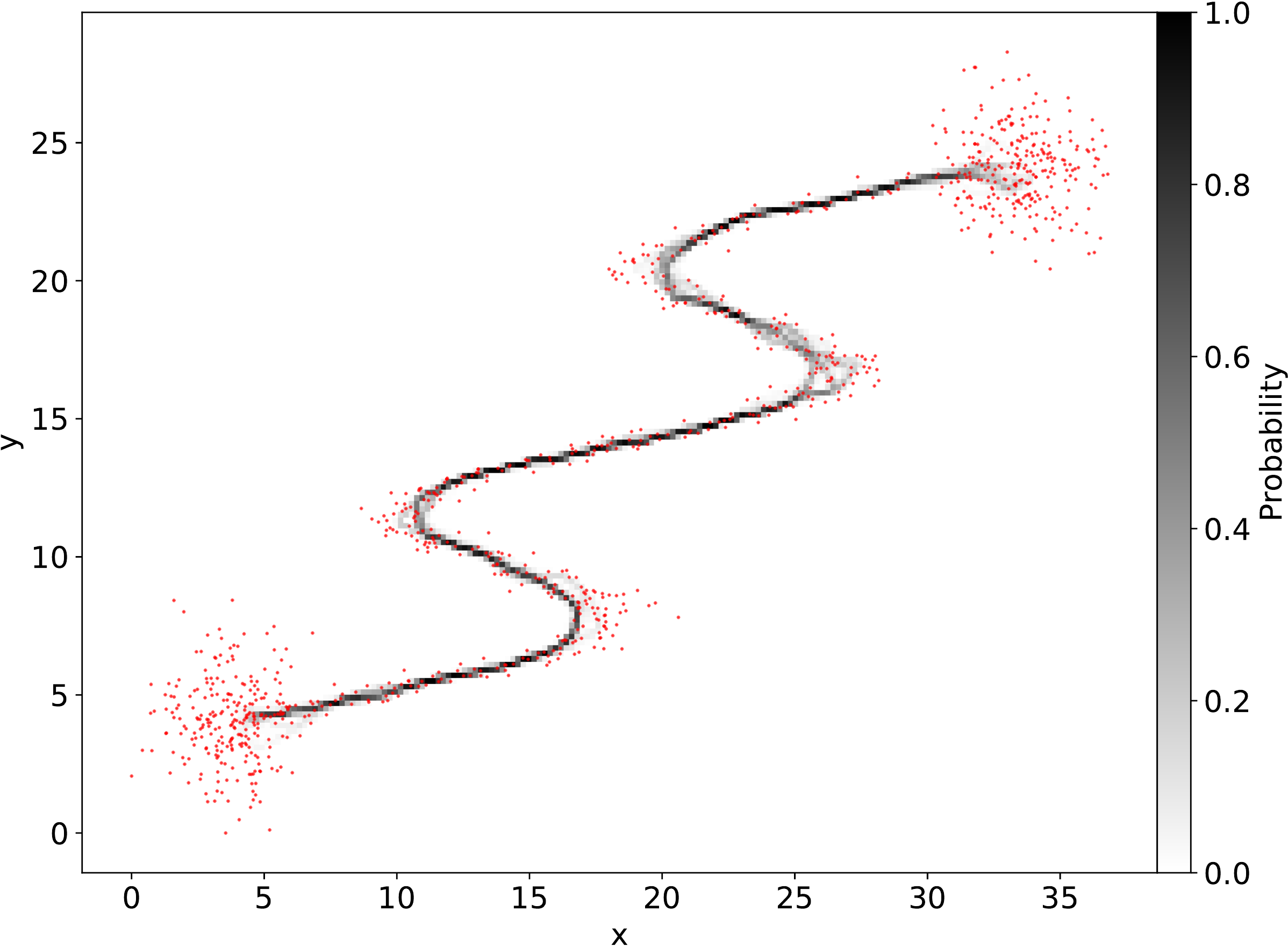}
        \end{subfigure}
        \caption{Probability map obtained from \autoref{Algo:Frequency_map} and eq. \eqref{eq:FreqMap} with $B=200$ and $N_B = 0.75N$. Red dots are input data points overplotted on the probability map.}
        \label{fig:Map_toyModel}
    \end{figure}

    \autoref{fig:Map_toyModel} shows a probability map obtained from the toy dataset in which the intensity of each pixel corresponds to the frequency that an edge of the MST crossed it. This way, we quantify the reliability of the various paths in the input domain. In practice, to build $\boldsymbol{I}(x)$, we use both the projected points $\boldsymbol{F}_b$ and the set of edges linking vertices encoded in $\mathcal{B}_b^R$ that contains information on the paths used and consequently should be taken into account in the final distribution. Edges are thus sampled and counted in the computation of $\boldsymbol{H}_b$ for Eq. \eqref{eq:FreqMap}. In what follows, we may refer to a quantity called the superlevel set of those maps defined as $\boldsymbol{\Gamma}_p(\boldsymbol{I}) = \{x \mid \boldsymbol{I}(x) \geq p\}$.
    Those sets are used to threshold the probability maps and keep only regions with a probability higher than $p$.
    
\section{Choice of T-ReX parameters} \label{Parameters}
    In \autoref{tab:Parameters}, we summarise the parameters of the algorithm together with their roles. We also give the baseline values further used in our study. As we are dealing with simulations of the Cosmic Web, we fix the cut-off level to a low value $l = 4$ and look for $B=100$ regularised minimum spanning trees using uniformly at random $75\%$ of the dataset for each sample. However, each of these parameters has a different and specific impact on the detection of the pattern that we discuss below.
    
    \begin{table*}
        \centering
        \caption{Parameters implied in the procedure and baseline values used in the presented results.}
        \label{tab:Parameters}
        
        \smallskip
        
        \begin{tabular}{cll}
            \toprule
            Parameter & Role & Used values \\
            \midrule
            $\lambda$ & Elastic constraint on centroids & $1$\\
            $\sigma^2$ & Spatial extension of Gaussian kernels & eq. \eqref{eq:Silverman}, $A_0 = 0.1$\\
            $l$ & The cut-off level to prune MST & $4$\\
            $B$ & Number of bootstrap samples & $100$\\
            $N_B$ & Size of bootstrap samples & $0.75 \, N$\\
            \bottomrule
        \end{tabular}
    \end{table*}
    
    \subsection{Elastic constraint $\lambda$}  \label{lambda_effect}
\begin{figure}
    \centering
    \begin{subfigure}
        \centering
        \includegraphics[width=1\linewidth]{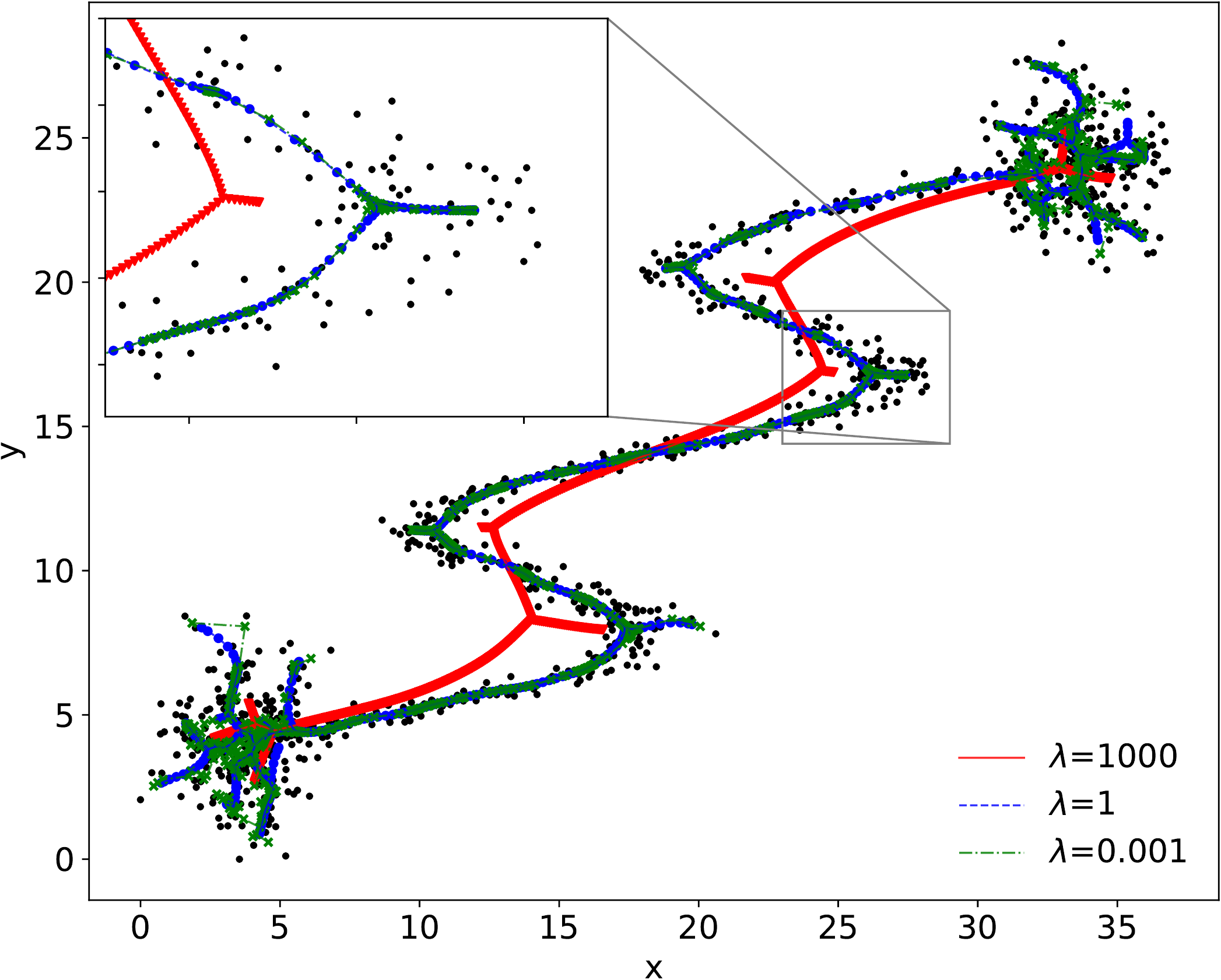}
    \end{subfigure}
    \caption{Effect of the $\lambda$ parameter on the regularised MST by fixing $\sigma=1$. Black dots are data points while red, blue and green lines are RMST with respectively $\lambda = \{1000, 1, 0.001\}$. Projected points are also represented, respectively by triangles, dots and crosses. We note that curves for $\lambda=1$ and $\lambda=0.001$ are almost superimposed.}
    \label{fig:Effect_lambda}
    \end{figure}

    As mentioned in \sectRefShort{RGMM}, $\lambda$ is a regularisation parameter acting like a trade-off between a set of centroids minimizing the data reconstruction error and the strength of the smooth tree topology we enforced.
    Hence, we understand that the larger $\lambda$, the more important the second part of eq. \eqref{RGMM_Mstep}, leading to a shorter and smoother tree, as seen on \figRefShort{fig:Effect_lambda}. $\lambda$ can be seen as a soft-constraint on the total length of the tree, a high value leading to a tree representation that has short extensions and projected points are more uniformly distributed over the tree.
    Given the definition of $\lambda$ in \sectRefShort{RGMM}, it is also the ratio between both variances of Gaussian kernels we used, one for the data fitting term and the other for the prior on centroids to introduce the elastic regularization term. Choosing $\lambda = 1$ thus induces that the two kernels have the same variance. When dealing with outliers or highly noisy datasets, $\lambda$ also helps increasing the robustness and maintains the tree structure in the desired regions without extending in noisy and underdense regions.
    
    \cite{Mao2015} proposed to tune $\lambda$ using the gap statistics, originally presented by \cite{Tibshirani2001} to choose the number of clusters in the K-means algorithm. This method requires several runs of the
    \autoref{Algo:PPT} with a range of $\lambda$ which can be very costly when dealing with large datasets. We hence choose to fix $\lambda = 1$ in our runs, leading to satisfactory results for a well chosen $\sigma$.
    
    \subsection{Spatial extension of Gaussian clusters $\sigma^2$}
\begin{figure}
    \centering
    \begin{subfigure}
        \centering
        \includegraphics[width=1\linewidth]{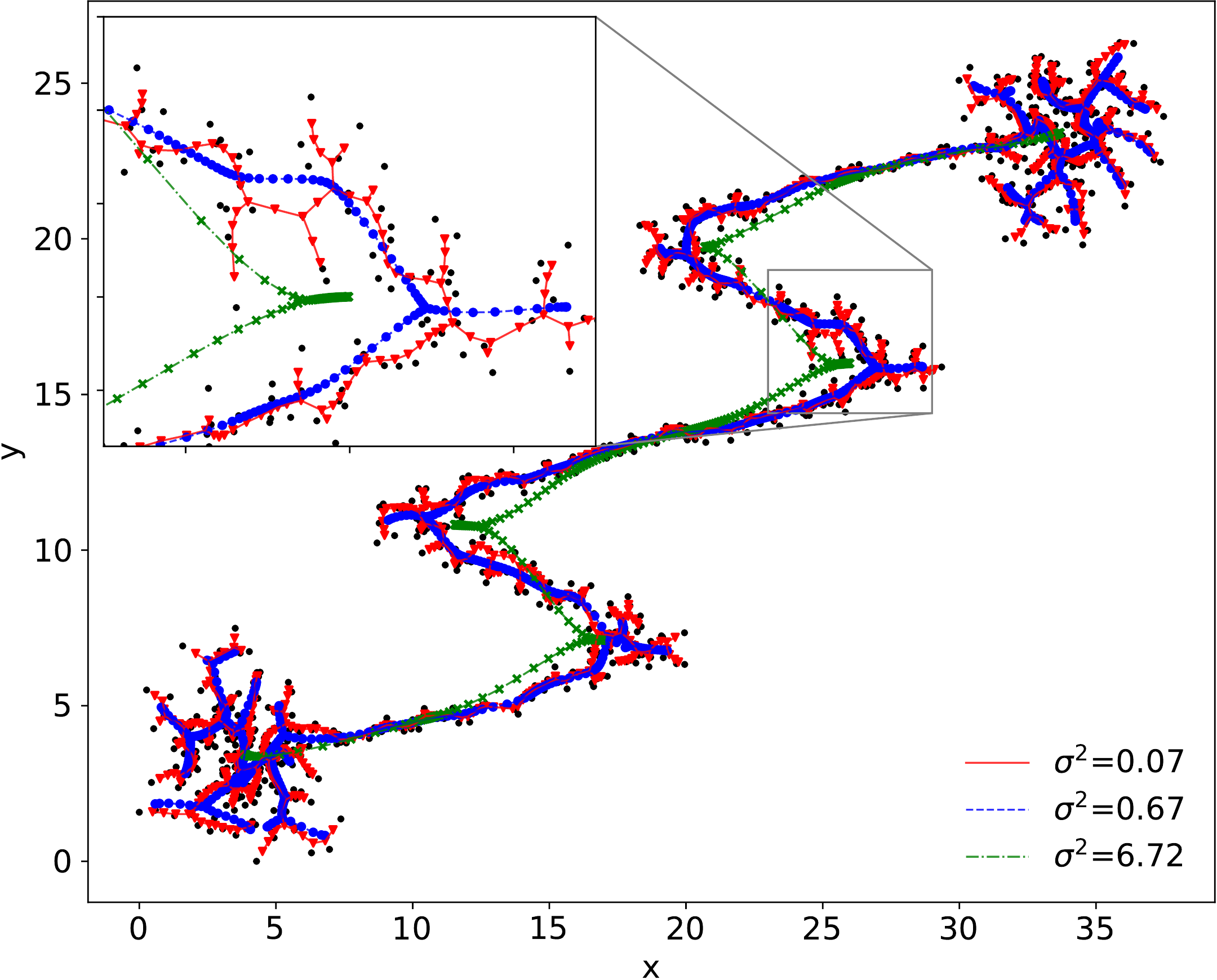}
    \end{subfigure}
    \caption{Effect of the $\sigma$ parameter on the regularised MST by fixing $\lambda=1$. Black dots are data points while red, blue and green lines are RMST with respectively $\sigma^2 = \{\frac{\sigma_s^2}{10}, \sigma_s^2, 10\sigma_s^2\}$ (see eq. \eqref{eq:Silverman}). Corresponding projected points are also represented, respectively, by triangles, dots and crosses.}
    \label{fig:Effect_sigma}
    \end{figure}

    The parameter $\sigma^2$ corresponds to the variance of Gaussian clusters used to compute the assignment matrix $\boldsymbol{P}$ in \autoref{Algo:PPT}. It ensures the local smoothness of the graph by allowing a soft partitioning of the input data points into centroids.
    Thus, $\sigma$ represents the spatial extension of each cluster and the higher it is, the more data points will be affiliated to a specific node of the resulting graph leading to a coarser representation. This trend is illustrated in \figRefShort{fig:Effect_sigma}, which shows several regularised MST obtained by fixing $\lambda = 1$ and varying $\sigma$. As $\sigma$ increases, centroids tend to be aligned and they describe a coarser shape of the underlying structure, biasing the estimate. Intuitively, $\sigma$ should represent the thickness of a typical filament so that centroids are fitting the distribution well.
    
    To automatically tune this parameter from the data, we follow the recommendation of \cite{Chen2015}, who  investigated the choice of such a parameter in the SCMS algorithm. We thus chose $\sigma$ using a modified version of the Silverman's rule \citep{Silverman1986}:
    \begin{equation} \label{eq:Silverman}
        \sigma_\text{s} = A_0 \Big(N (d+2) \Big)^{\frac{-1}{d+4}} \sigma_{\text{min}},
    \end{equation}
    where $A_0$ is a constant, $N$ is the number of data points, $d$ is the dimension of the data and $\sigma_{\text{min}}$ is the minimum standard deviation over all directions. Taking $A_0 = 1$ leads to the Silverman's rule and is the optimal estimate for an underlying Gaussian distribution. As argued by \cite{Chen2015}, when the data are not Gaussian anymore, $A_0$ should be optimised as a free parameter.
    In our experiments, when the parameter is not explicitly defined, we adopt the baseline value of \autoref{tab:Parameters}, namely $A_0 = 0.1$, a rather low value so that the estimated trees keep some small scales variations. When $A_0$ increases, the smoothing scale also increases and a coarser filamentary pattern is described.
    
    Although we considered a fixed isotropic and identical covariance matrix for all clusters, it is noteworthy that the formalism initially presented in \sectRefShort{RGMM} is more general. We could consider a specific covariance for each cluster, initialise it with the rule of Eq. \eqref{eq:Silverman} and adapt it automatically from the data. EM computation can indeed auto-adjust this estimate at each iteration by considering $\theta = \{f_1, \ldots, f_k, \Sigma_1, \ldots, \Sigma_k\}$ and then maximising the lower bound of the log-likelihood not only over $f_k$ but also with respect to $\Sigma_k$ in the M-step. This solution has, however, an additional computational cost and can lead each Gaussian cluster to be housed in a specific data point when $K$ is close to $N$. It did not sufficiently improved the results in our cosmological application to consider it but could be included in future works. The current choice, hence, restricts the range of scales that can be described by the Gaussian clusters, implying that broad structures in which the extension is way above $\sigma$ will not collapse into a single ridge passing in the middle of the structure in the resulting graph.
    
    \begin{figure*}
        \centering
        \subfigure{\includegraphics[scale=0.35]{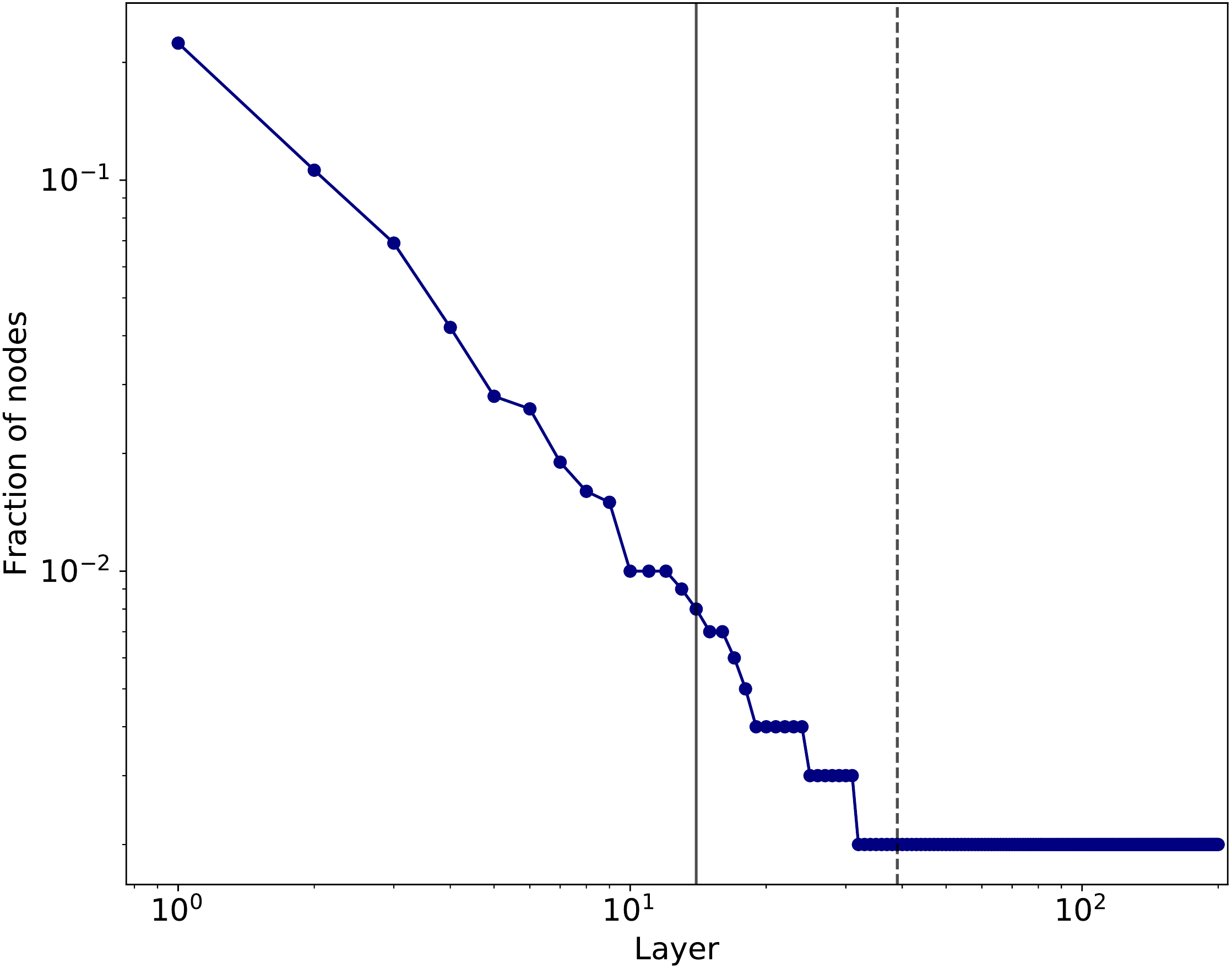}} \quad
        \subfigure{\includegraphics[scale=0.35]{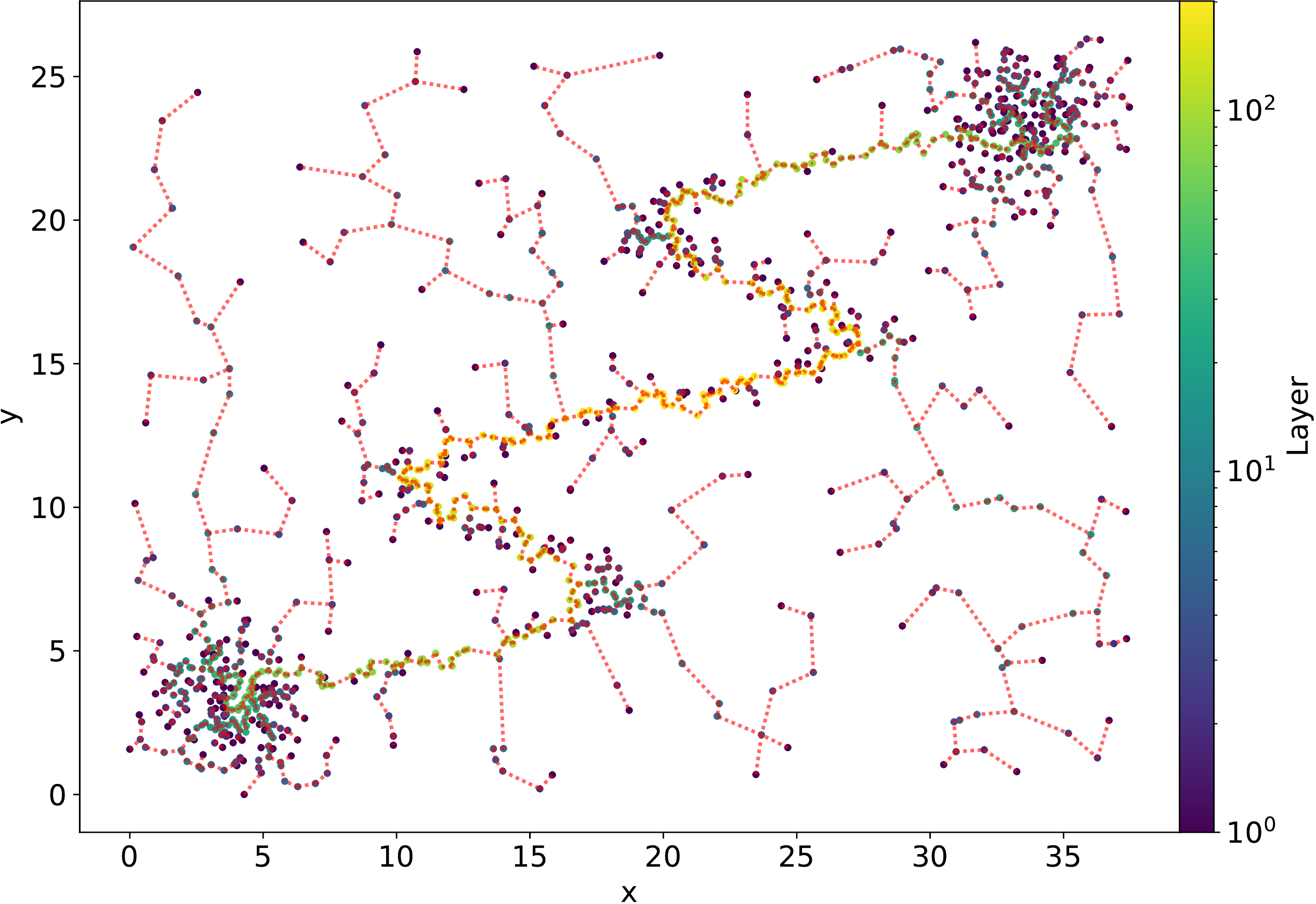}}
    
        \caption{\textit{(left)} Onion spectrum of the tree structure. Vertical lines correspond to $l=14$ (solid) and $l=39$ (dashed) discussed in \sectRefShort{Pruning}. \textit{(right)} Layer value of each datapoint. Red dashed line is the MST and dots are data points from a noisy version of the toy dataset (obtained by adding $25$\% uniform noise in the bounding box).}
        \label{fig:Onion}
    \end{figure*}
    
    \subsection{Pruning level $l$} \label{Pruning}
    As explained in \sectRefShort{Pruning}, the pruning acts as a denoising operation but it also helps reducing the number of kernels to span the point cloud. A high cut-off level removes a large number of nodes at the extremity of all branches revealing only the core of the tree structure while a lower value allows branches to have long extensions reaching even nodes in empty regions. The choice of $l$ can hence lead to different tree representations of a noisy dataset.
    
    To choose the value of $l$, we rely on the work of \cite{Hebert-Dufresne2016} who introduced the onion decomposition for graphs. The idea is to attribute to each node a layer in terms of depth in the network allowing to define a center and a periphery. The left panel of \figRefShort{fig:Onion} shows the onion spectrum of the noisy toy dataset and illustrates the points of the noisy dataset colored by their layers. 
    The power-law decay in the first part of the spectrum can be interpreted as the removal of all short branches (in number of nodes). A constant level in the onion spectrum means that we are iteratively removing the same amount of nodes in the network at each iteration and thus that the tree structure is 'stable' in terms of number of branches.
    Using as cut-off level the beginning value of the last constant level in the onion spectrum ($l=39$ on left of \figRefShort{fig:Onion}) would lead to keeping only the core of the tree structure with a single branch (the longest one in the initial tree). However, this is a very conservative solution and doing so in real datasets would lead to miss some end parts of filaments or peripheral structures.   
    
\begin{figure}
        \centering
        \begin{subfigure}
            \centering
            \includegraphics[width=1\linewidth]{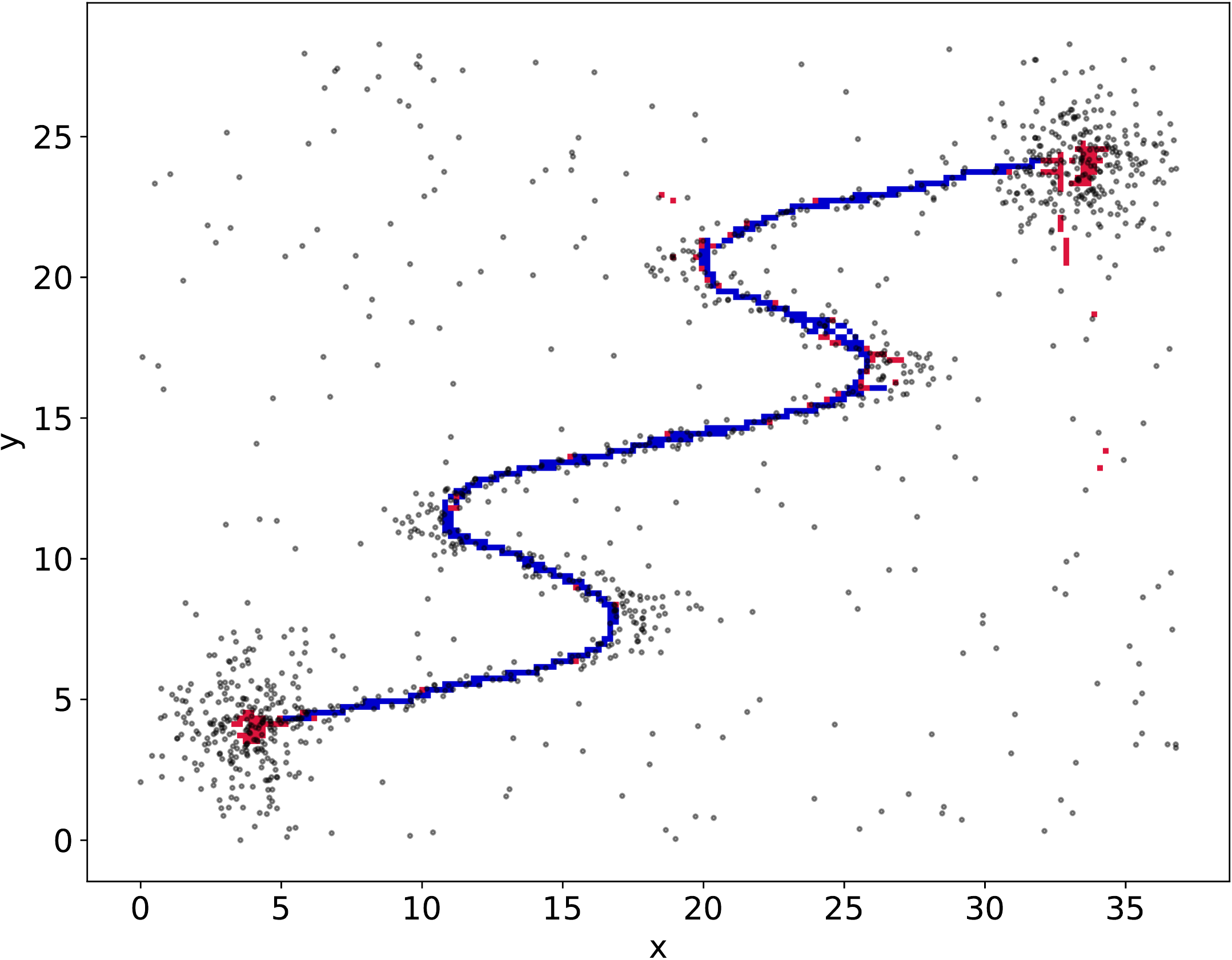}
        \end{subfigure}
        
        \caption{Superlevel sets $\boldsymbol{\Gamma}_{0.25}(\boldsymbol{I})$ for two pruning levels on the noisy toy dataset: $l=14$, a too low cut-off and $l=39$, an adequate value. Blue pixels are regions where both sets are overlapping while red ones show regions highlighted by the $l=14$ version but not by the $l=39$ one and are mostly found in the background noise of the pattern.}
        \label{fig:Map_pruning}
    \end{figure}

    A threshold $l$ that is too low can bring out spurious detections of the underlying pattern for a realization of a tree. However, this effect should be mitigated by: i) the $\lambda$ parameter which also helps reducing the length of branches in noise and outliers, as discussed in \sectRefShort{lambda_effect} and ii) the bootstrap step where those detections will have a low occurrence as illustrated in the superlevel sets of \figRefShort{fig:Map_pruning}. For this reason, in what follows, we consider a rather low value for the pruning parameter, namely $l=4$. In the case of simulated cosmological datasets, this parameter only helps to remove data points that are located in empty or low dense regions since, for a well chosen value of $\sigma$, \autoref{Algo:PPT} is robust to noise encountered around the ridges.
    
    \subsection{Number and size of the bootstrap samples}
    Both the number and size of the replicated samples, $B$ and $N_B$ respectively, are related to the probability map. Above a minimum value, the parameter $B$ has almost no effect on the estimate for a fixed $N_B$. The main idea to explain this phenomenon is that, for a fixed size $N_B$, there is only a limited number of different possible paths with high probability. Even though a higher value for $B$ can highlight some new paths, they will have a very low occurrence.

    $N_B$ affects the map in a more important way. A low size value induces more possible paths to cross and thus more variability in the resulting map while a size close to the initial one $N$ ($0.90 N$ for instance) allows for only local modifications of the highlighted paths; hence, it is more conservative. Choosing a low value can thus lead to more spurious path detection.


\section{Results: application to cosmological datasets} \label{Results}

    In this section, we apply T-ReX with the baseline parameters of \autoref{tab:Parameters} on the 2D and 3D cosmological datasets described in \sectRefShort{Data}. The slice of the 2D subhalo distribution corresponds to the data points in \figRefShort{fig:Illustris3} which represent a projected slice of $5$Mpc/h depth. The 3D distribution of halos is built from a $200^3$Mpc/h Gadget-2 simulation used in \cite{Libeskind2017}.
    
    \subsection{Filamentary structure in a 2D subhalo distribution} \label{2D}
    
\begin{figure}
        \begin{subfigure}
            \centering
            \includegraphics[width=1\linewidth]{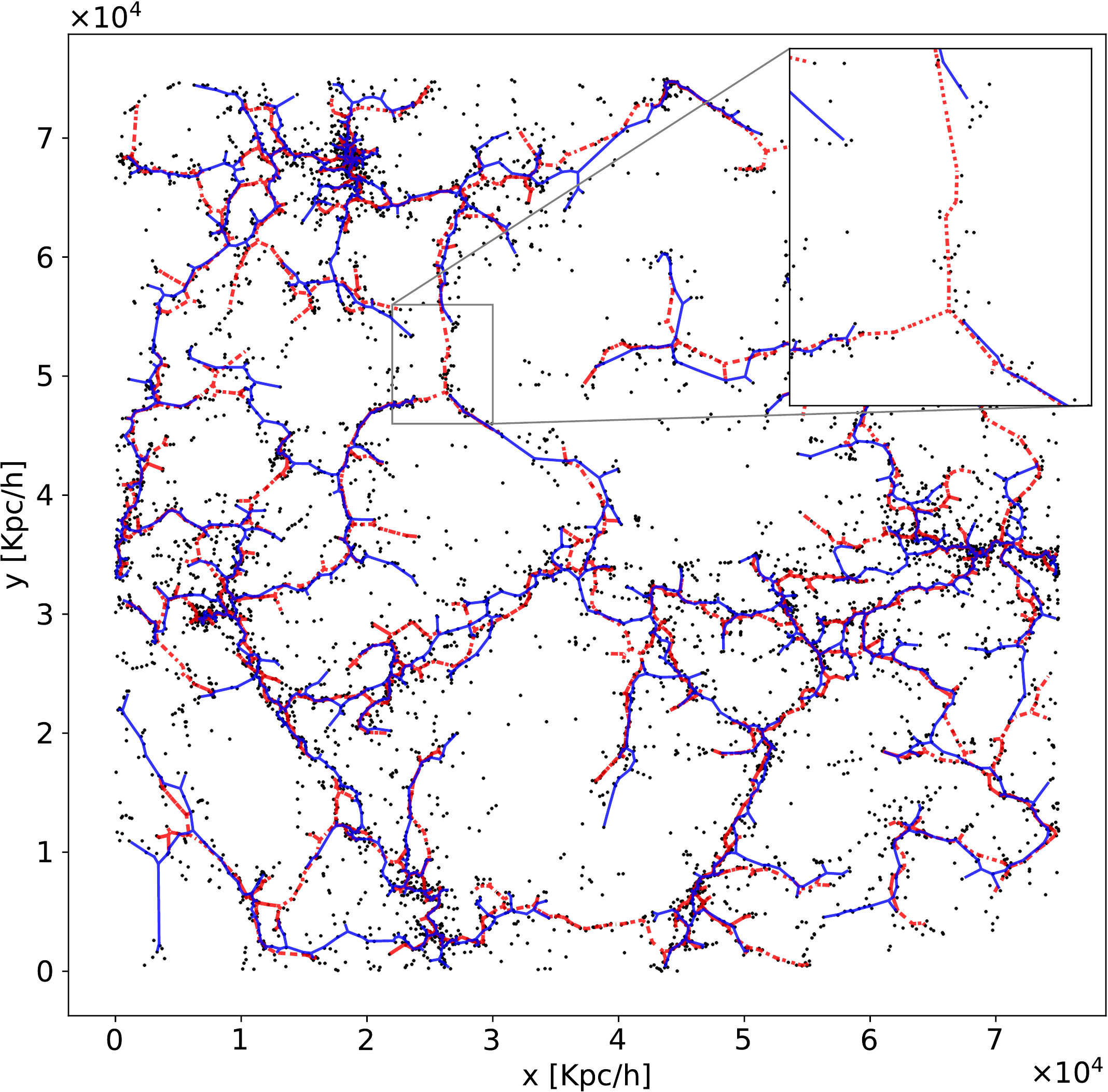}
        \end{subfigure}
        
        \caption{Two realisations (solid blue line and dotted red line) of RMST (\autoref{Algo:PPT}) with $75\%$ of the dataset picked randomly and uniformly with the parameters of \autoref{tab:Parameters}. Black dots are subhalos from the Illustris-3 simulation.}
        \label{fig:RMST_Illustris}
    \end{figure}

    \autoref{fig:RMST_Illustris} shows two realisations of a regularised minimum spanning tree (see \autoref{Algo:PPT}) with over $75\%$ of the data points picked randomly and uniformly. Firstly, it is interesting to see that each RMST is standing in regions that would naturally be called ridges or filaments in the distribution of galaxies, that is, elongated structures connecting high density regions together. Secondly, we can see that according to the the distribution of picked data points, different paths are taken for the core of the tree structure. The complementarity of the two realisations is highlighted in the zoomed region. Since the tree topology cannot include loops, the effect of disconnection is observable in this particular region where the solid blue realisation is not fully connecting the network. We note that such an effect is intensified by the pruning operation. Other realisations might not exhibit the disconnection in the same region as seen in the case of the dashed red line of \figRefShort{fig:RMST_Illustris}. This highlights the necessity and the interest of stacking several RMST to obtain a full characterisation of the cosmic network.
    
\begin{figure}
        \centering
        \begin{subfigure}
            \centering
            \includegraphics[width=1\linewidth]{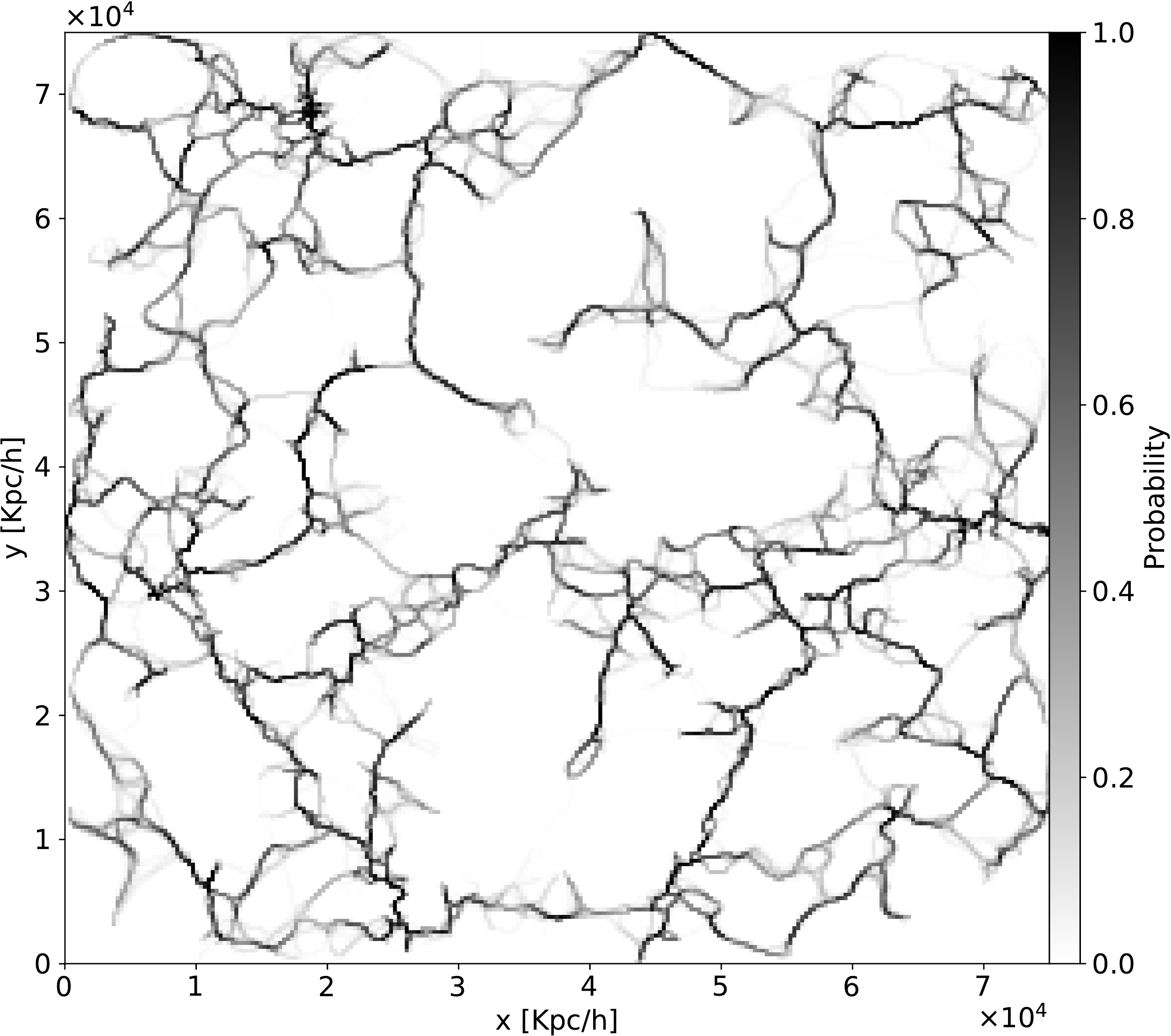}
        \end{subfigure}
        
        \begin{subfigure}
            \centering
            \includegraphics[width=1\linewidth]{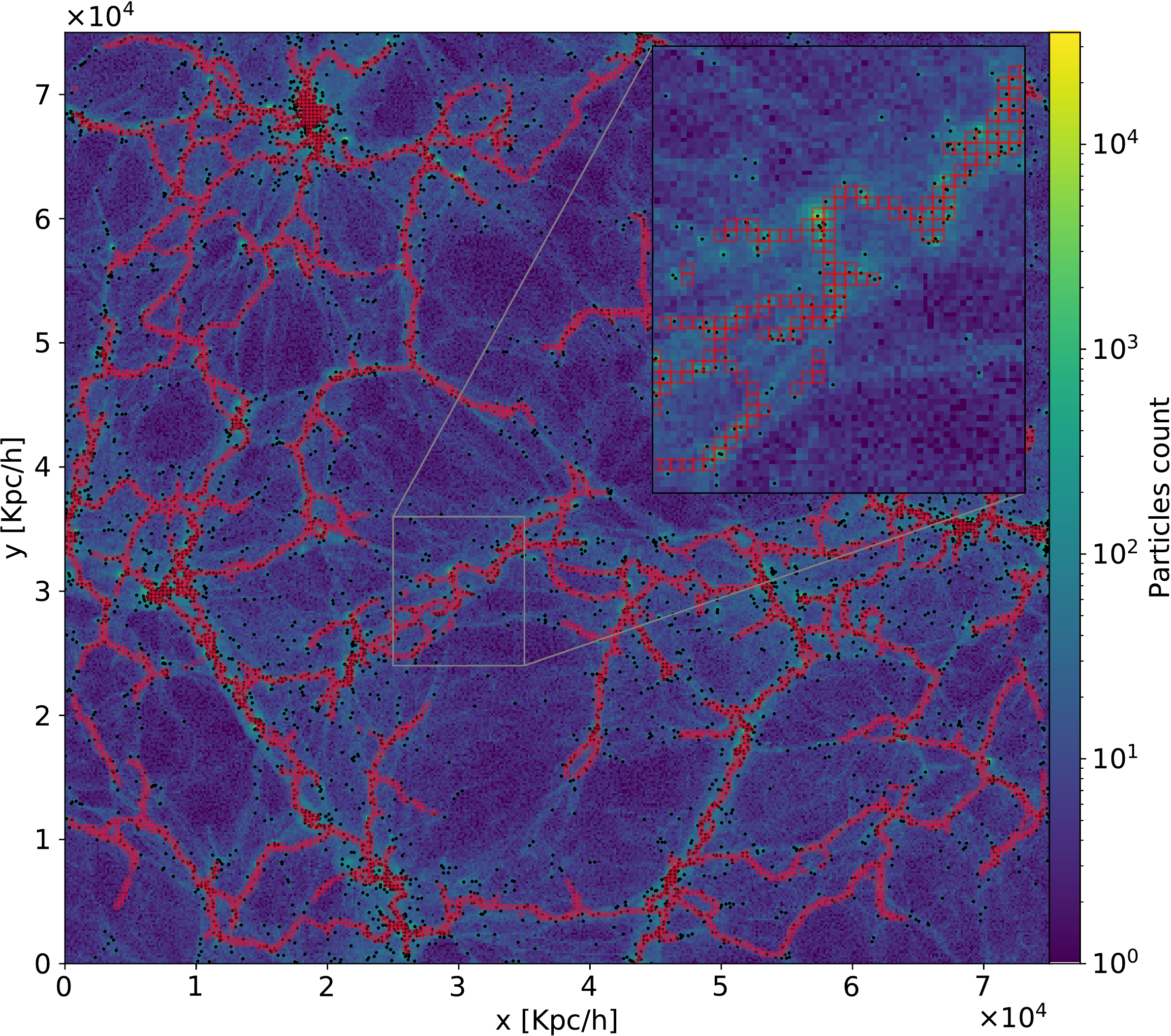}
        \end{subfigure}
        \caption{\textit{(top)} Probability map $\boldsymbol{I}$ obtained from subhalos displayed in \figRefShort{fig:Illustris3} with parameters described in \autoref{tab:Parameters}. The resolution of the probability map is $250$Kpc/h. \textit{(bottom)} Superlevel set $\boldsymbol{\Gamma}_{0.25}(\boldsymbol{I})$ (red squares) overplotted on the DM distribution together with subhalos (black dots).}
        \label{fig:Res_Illustris}
    \end{figure}

    \autoref{fig:Res_Illustris} shows a probability map obtained from $100$ realizations of RMST. We can see that the highly probable part of the map is fitting what one would expect for the underlying distribution while the overlap of the superlevel set $\boldsymbol{\Gamma}_{0.25}(\boldsymbol{I})$ with the DM distribution allows us to see that high probability paths (above $0.25$ in this case) are tracing the most prominent part of the network. It is worth noting that the agreement is particularly interesting given that the input of the algorithm are subhalos and not DM particles. The zoomed-in region emphasises that small scales are also recovered where high probability paths follow the ridge in the DM distribution.
    
    \subsubsection{Comparison with DisPerSE skeletons} \label{Disperse}
    DisPerSE \citep{DisperseTheory} is a publicly available\footnote{\url{http://www2.iap.fr/users/sousbie/web/html/indexd41d.html}} and widely used algorithm capable of detecting filaments and walls in a density field tracer, such as galaxy distribution. From this discrete set of particles, a continuous density field is estimated using the Delaunay tessellation field estimation. Based on the discrete Morse theory \citep{Forman1998}, DisPerSE first aims at identifying singularities (or critical points) in the field defined as positions where the gradient cancels and then uses the local morphology to classify those points in maxima, minima, and saddles using eigenvalues of the Hessian matrix.
    DisPerSE finally identifies filaments using the connectivity of critical points following the gradient lines in the density field. Persistent homology \citep{Edelsbrunner2002} is then used to remove insignificant parts of the pattern.
    
\begin{figure*}[h]
    \centering
    \setkeys{Gin}{width=0.4\textwidth}
    
    \hspace*{\fill}
        \centering
        \includegraphics[width=0.4\textwidth]{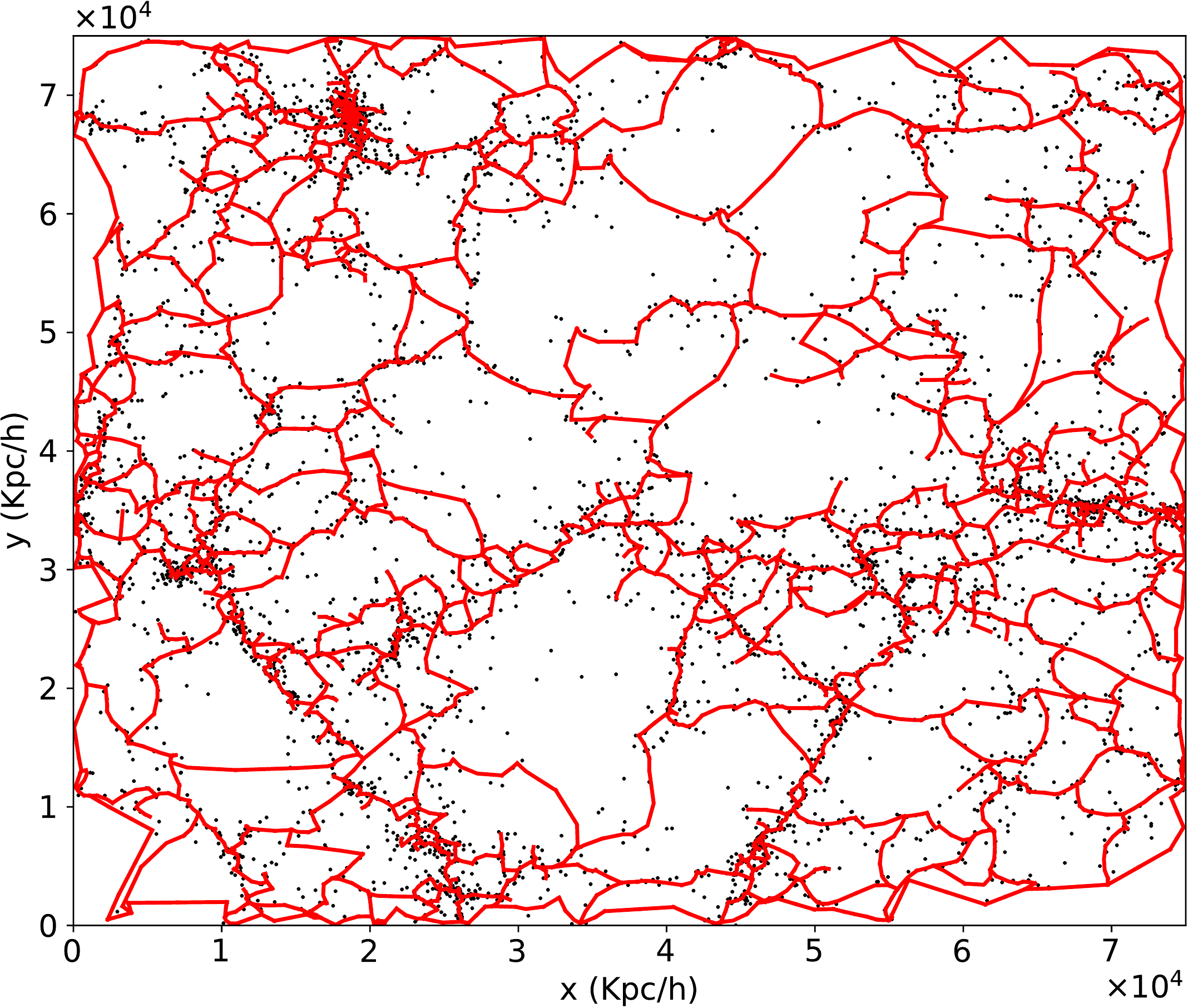}\hfill
        \includegraphics[width=0.4\textwidth]{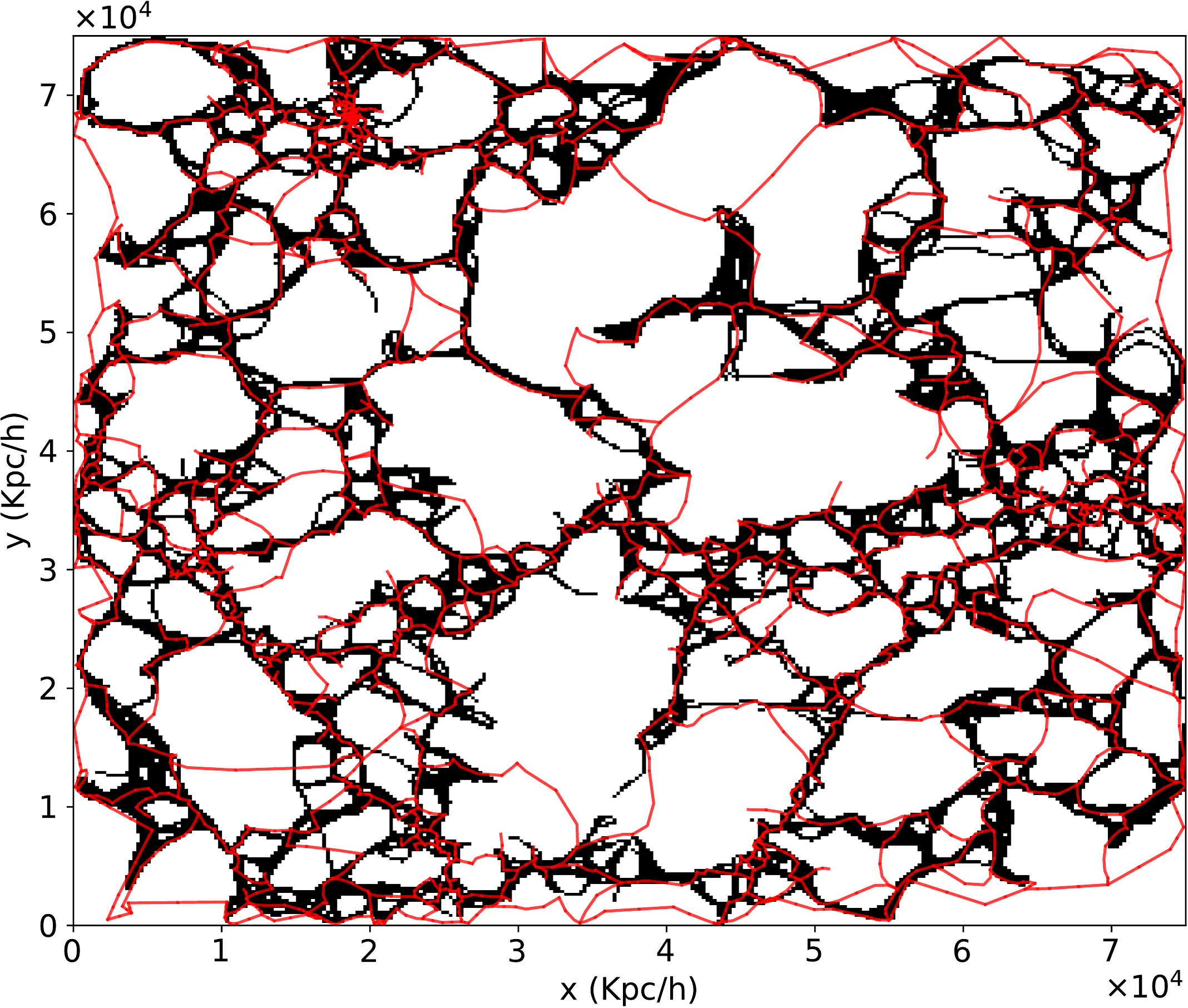}
    \hspace*{\fill}

    \hspace*{\fill}
        \centering
        \includegraphics[width=0.4\textwidth]{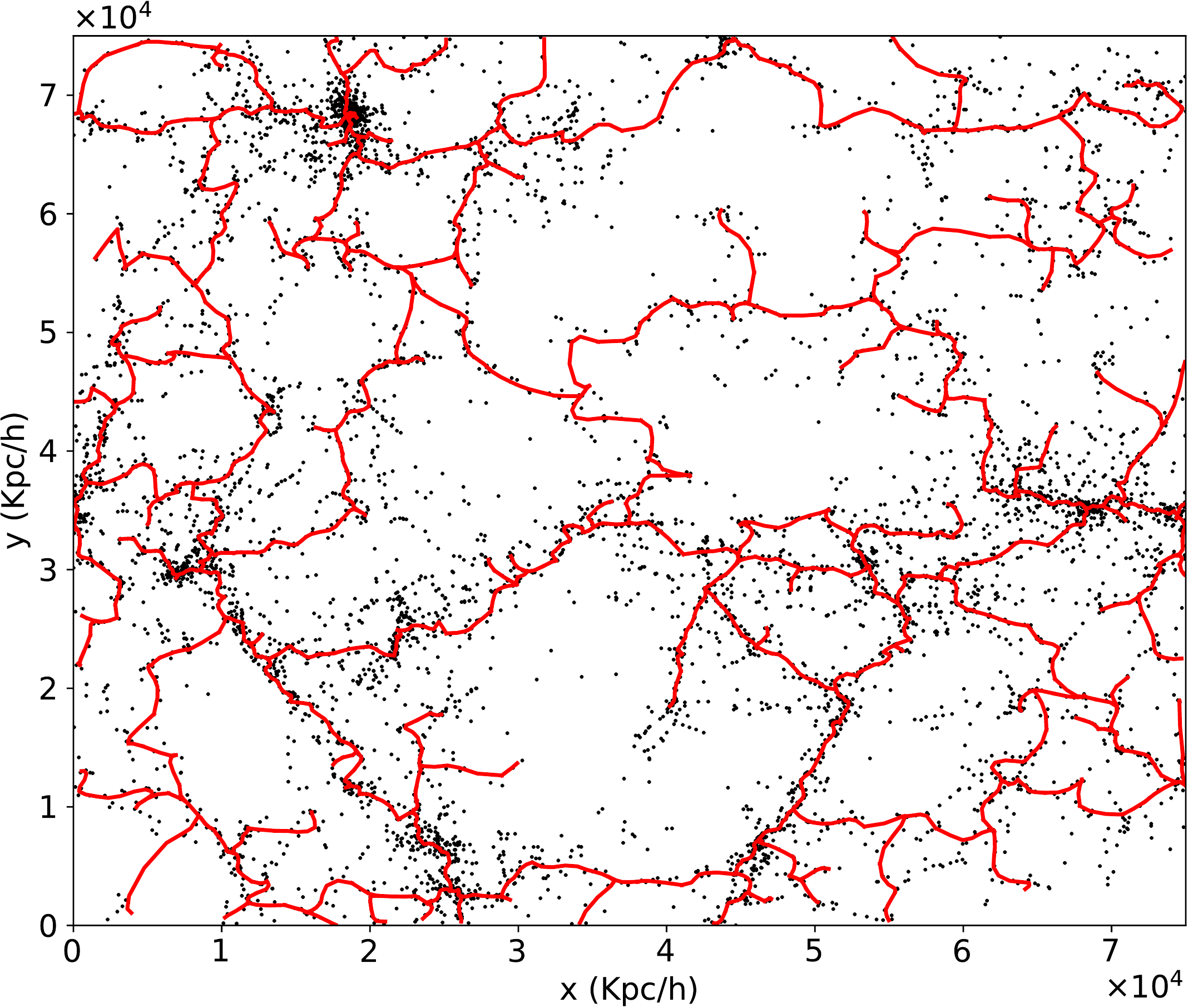}\hfill
        \includegraphics[width=0.4\textwidth]{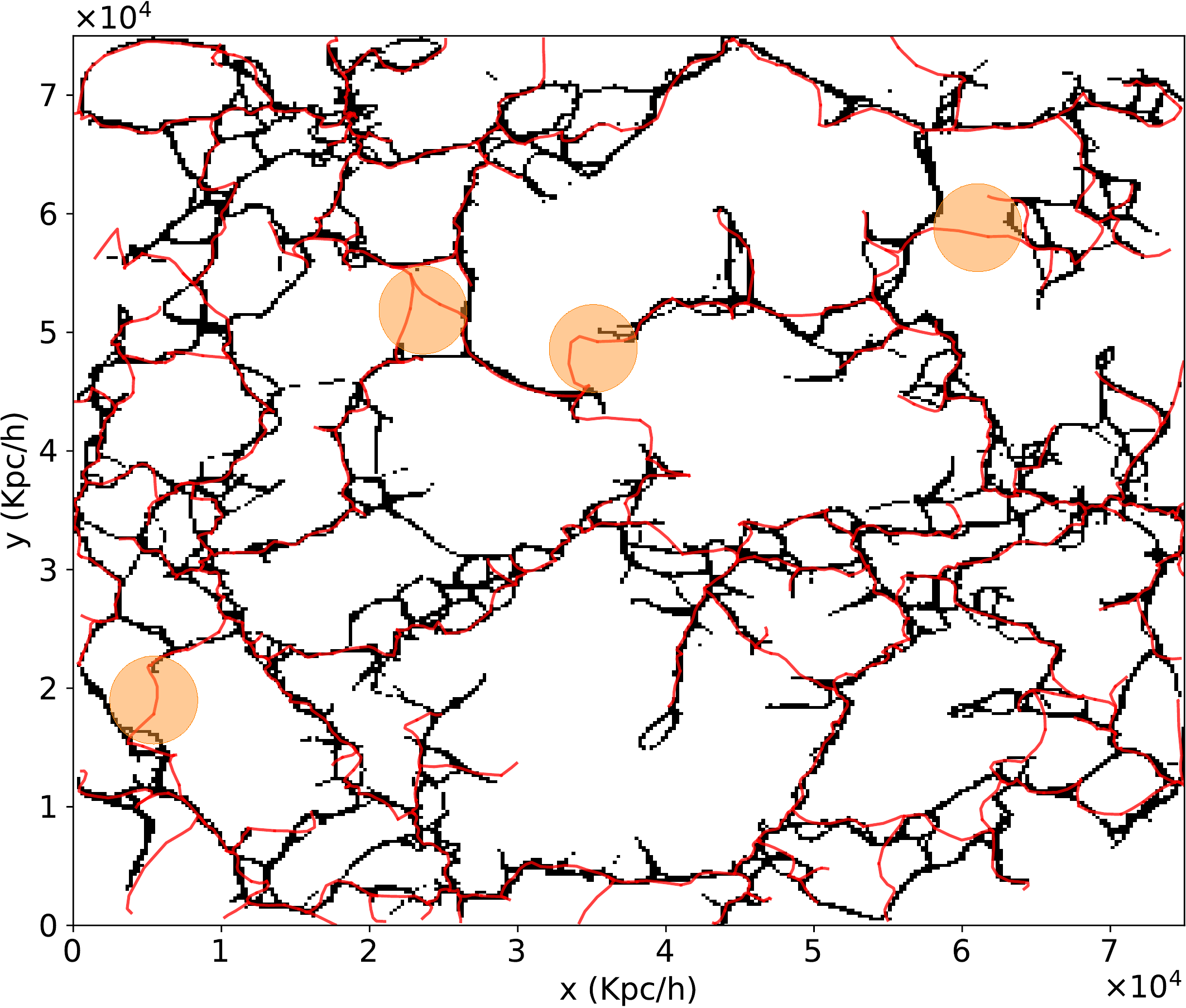}
    \hspace*{\fill}

    \hspace*{\fill}
        \centering
        \includegraphics[width=0.4\textwidth]{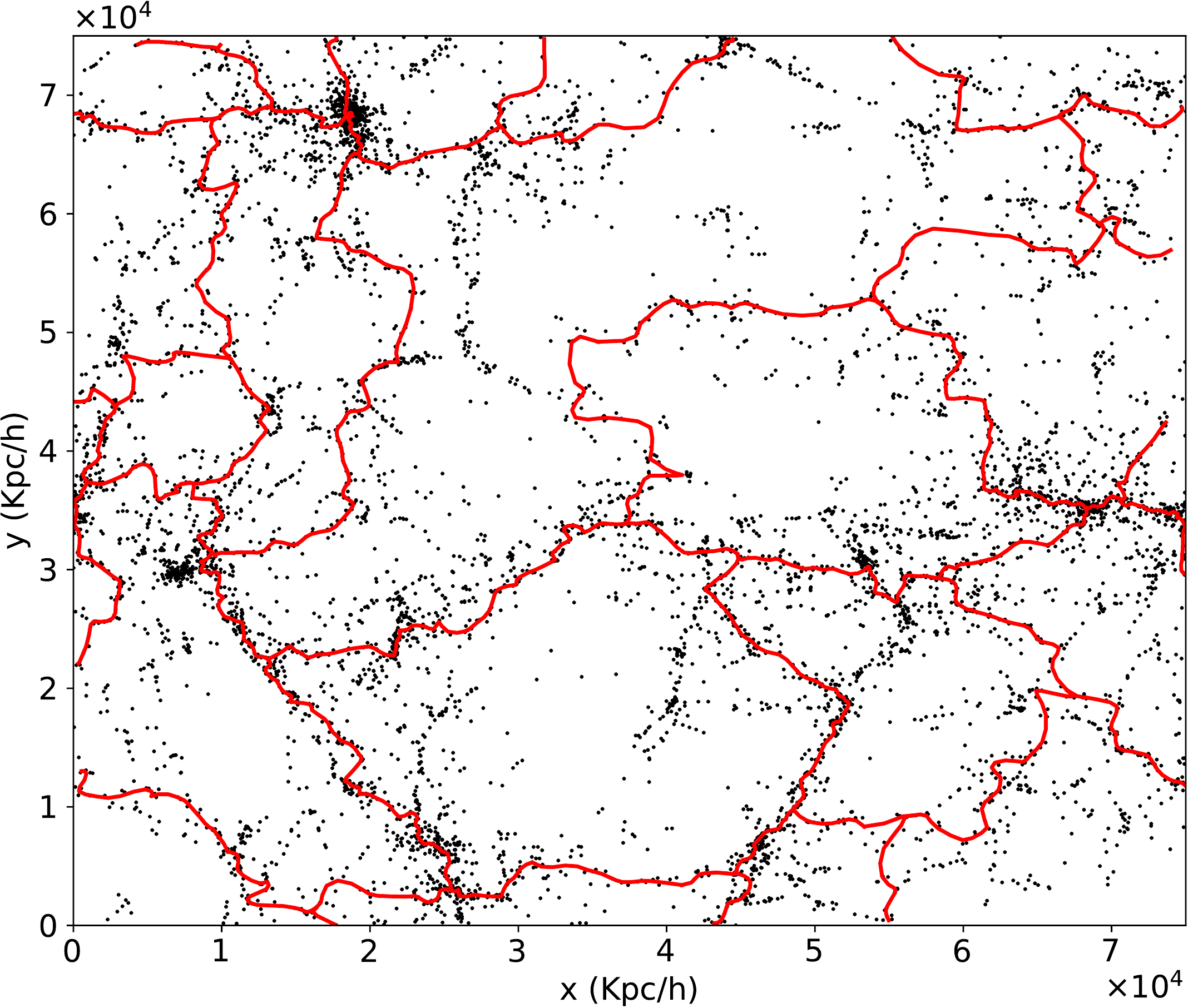}\hfill
        \includegraphics[width=0.4\textwidth]{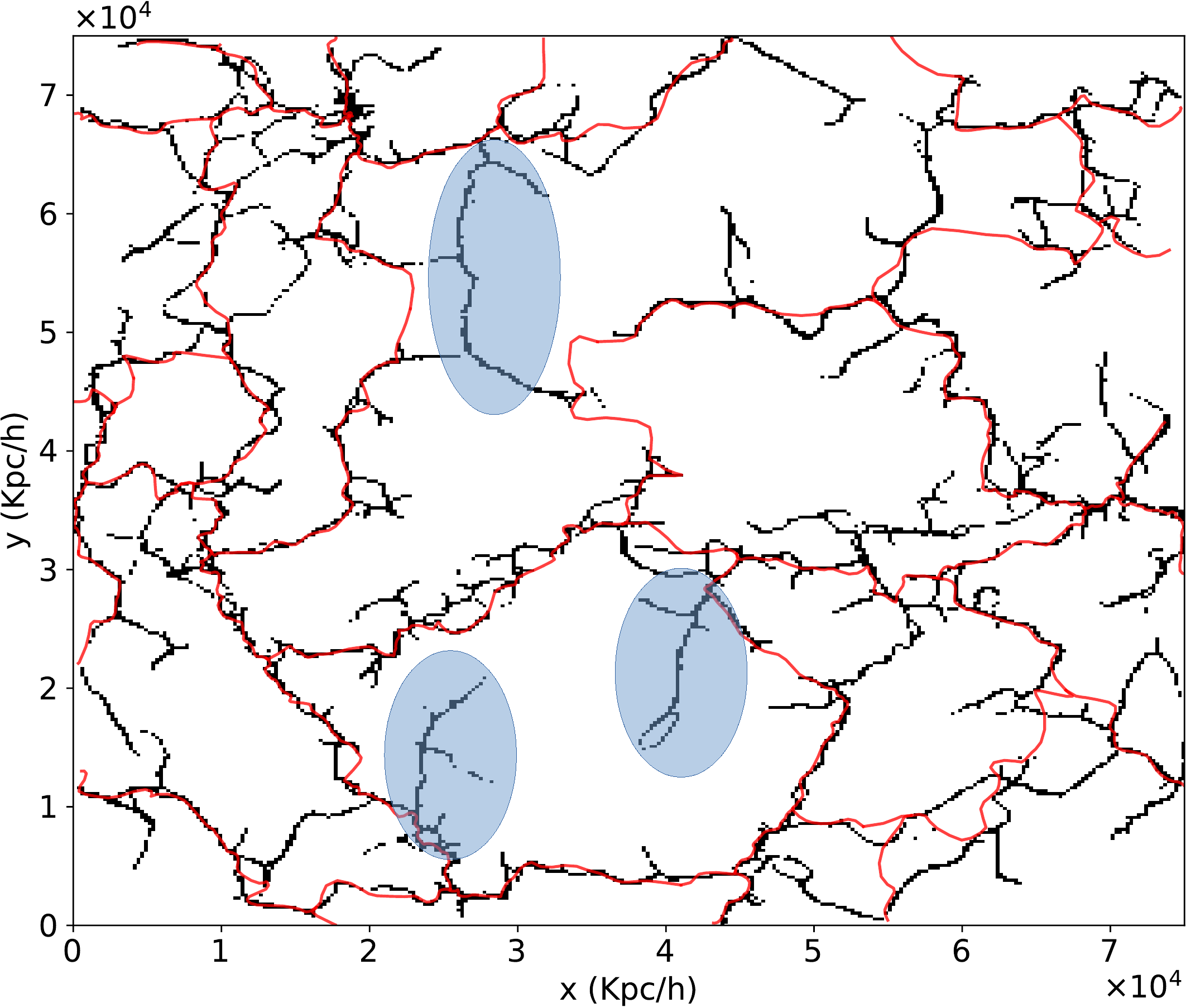}
    \hspace*{\fill}
    
    \medskip
    
    \caption{\textit{(left column)} Subhalos (black dots) and DisPerSE skeletons (red lines) with several significance levels (from top to bottom: $\sigma_p = 0, 2, 5$). \textit{(right column)} Superimposition of some thresholded probability maps obtained by T-ReX and DisPerSE skeletons (red lines) with several significance levels (from top to bottom: $\sigma_p = 0$ and $\boldsymbol{\Gamma}_{0.0}(\boldsymbol{I})$, $\sigma_p = 2$ and $\boldsymbol{\Gamma}_{0.1}(\boldsymbol{I})$, $\sigma_p = 5$ and $\boldsymbol{\Gamma}_{0.25}(\boldsymbol{I})$). Resolution of the maps provided by T-ReX is $250$Kpc/h. Shaded blue and orange areas highlight some differences between results discussed in \sectRefShort{Disperse}.}
    \label{fig:Disperse_output}
    \end{figure*}

    \autoref{fig:Disperse_output} shows both T-ReX and DisPerSE results obtained on the Illustris slice with several probability thresholds for the former and several persistence levels for the latter. At fixed density smoothing (here 1), the DisPerSE skeletons show the best overlap with our un-tresholded probability map for a persistence $\sigma_p=0$, where the two methods agree for most of the filamentary structure. The boundary effects observed in the DisPerSE skeleton at low $\sigma_p$ disappear with increasing persistence. The good agreement between high probability paths provided by T-ReX and the DisPerSE skeleton remains with increasing persistence levels and probability thresholds as shown by the overlap of DisPerSE and T-ReX skeletons (right column of \figRefShort{fig:Disperse_output}). It should be emphasised that there is no direct transposition of the persistence threshold in DisPerSE into the probability threshold in T-ReX. The present choice of threshold parameters is hence arbitrary and only serves illustration purposes.
    
    Although the two algorithms have very different definitions for what they both call filamentary pattern, it is reassuring to see that they are recovering similar structures. However, it is not surprising to observe some disagreement on specific filaments (see orange shaded regions in \figRefShort{fig:Disperse_output}). Since the pattern identified by T-ReX is obtained by minimising a global criterion, some paths identified by DisPerSE are not relevant for minimising the total distance and, thus, they do not appear as possible paths in any of the realisations.
    When comparing two conservative cases, namely, $\boldsymbol{\Gamma}_{0.25}(\boldsymbol{I})$ and the $5\sigma$ DisPerSE persistence skeleton (lowest right panel of \figRefShort{fig:Disperse_output}), we see that the T-ReX pattern preserves more small-scale structures and provides some paths that seems coherent with the subhalo distribution but which are not identified with the chosen parameters for the DisPerSE output (see blue shaded regions in \figRefShort{fig:Disperse_output}).
    
    \subsubsection{Sparse data point distribution}    \label{Sparse}
\begin{figure*}[h]
        \centering
        \subfigure{\includegraphics[scale=0.45]{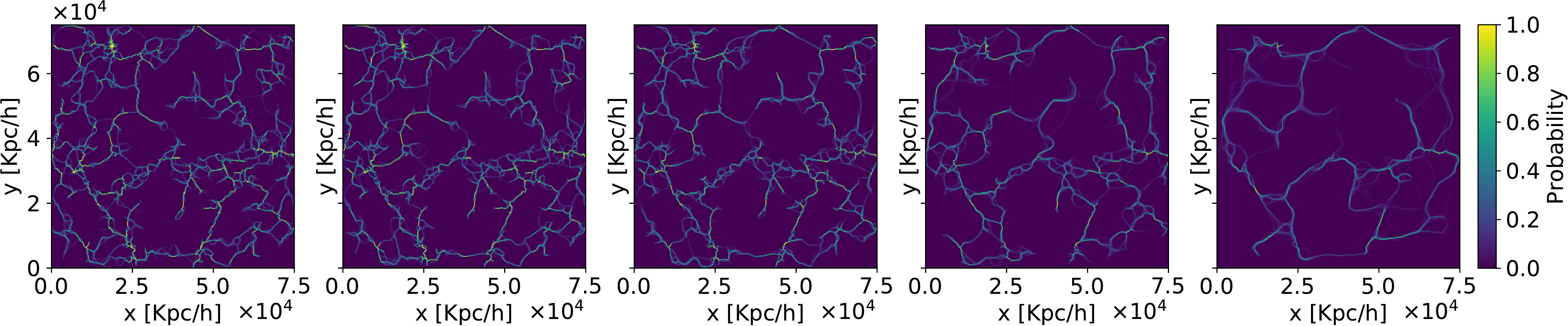}}
    
      \caption{Probability maps with increasing mass threshold $M^{\text{cut}}$. From left to right, $M^{\text{cut}} = \{0, 0.85, 1.35, 3.22, 11\}\times 10^{10} M_\odot/h$ corresponding, respectively, to $100\%$, $83\%$, $60\%$, $31\%,$ and $10\%$ of the total subhalos in the slice.}
      \label{fig:Sparse_freqMap}
    \end{figure*}

    In order to explore the robustness of the method against the datapoint density used for ridge detection, we reduce the number of subhalos in the initial dataset by keeping only those with a mass $M \geq M^{\text{cut}}$. In practice, we investigate how the original filamentary map is spatially close to the recovered ones when $M^{\text{cut}}$ varies. \autoref{fig:Sparse_freqMap} shows probability maps obtained for increasing values of $M^{\text{cut}}$ leading to sparser and sparser input ($100\%$, $83\%$, $60\%$, $31\%$ and $10\%$ of the initial subhalos in the slice respectively corresponding to $M^{\text{cut}} = \{0, 0.85, 1.35, 3.22, 11\}\times 10^{10} M_\odot/h$).
    Visually, probability maps show a nice stability, even when the sparsity is high: patterns are pretty much the same when we keep at least $60\%$ of the most massive objects hence recovering the essential part of the structure.
    
\begin{figure}
        \centering
        \includegraphics[width=1\linewidth]{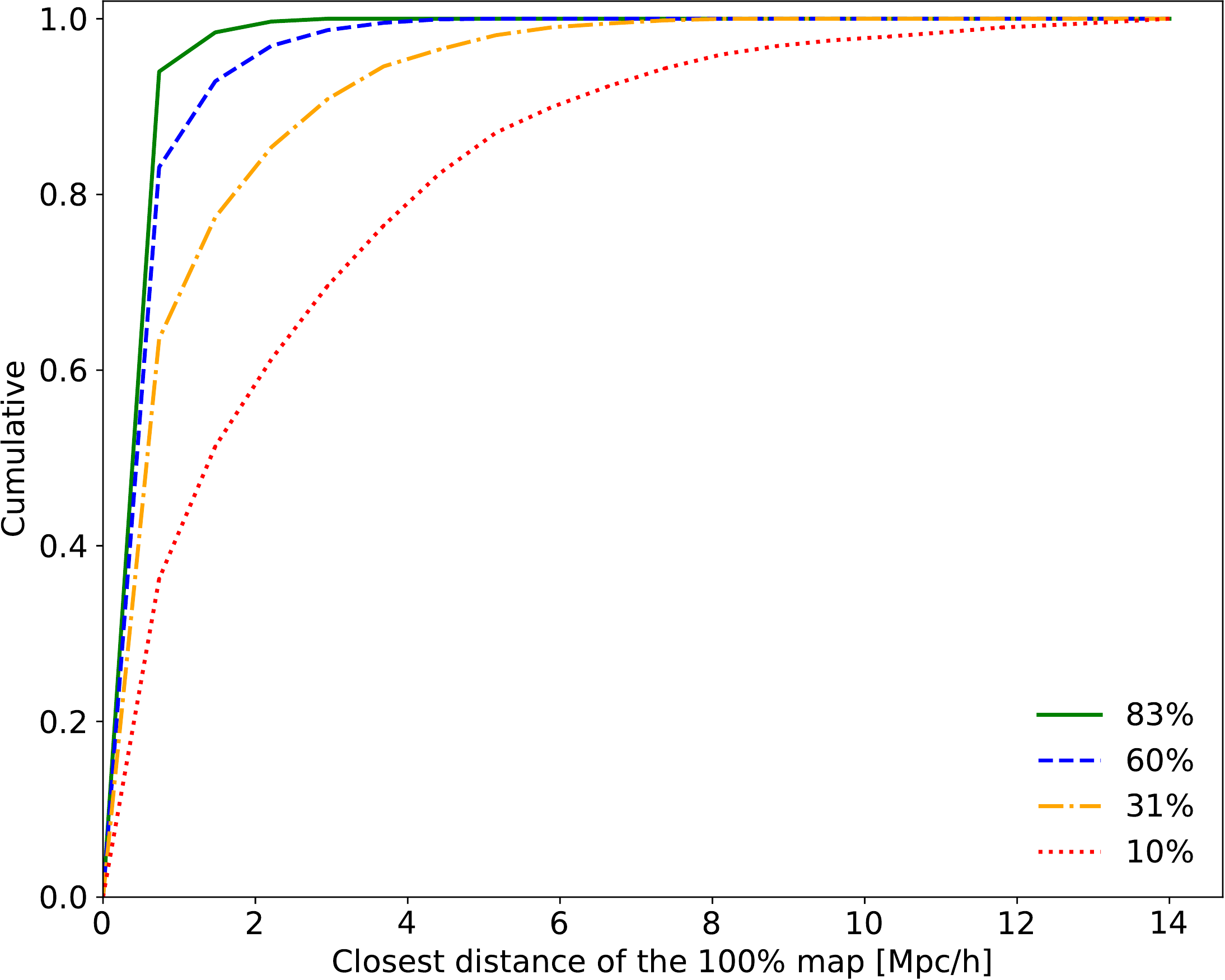}
        
        \medskip
        
        \caption{Cumulative distribution of distances $\{d_x^J\}$ (see \sectRefShort{Sparse}) between positions of the binary maps $\boldsymbol{\Gamma}_{0.25}(\boldsymbol{I}_{100})$ obtained with increasing mass threshold $M^{\text{cut}}$ to the one with $J\%$ of the data points. $M^{\text{cut}} = \{0.85, 1.35, 3.22, 11\}\times 10^{10} M_\odot/h$ leading respectively to $83\%$, $60\%$, $31\%$ and $10\%$ of the total number of subhalos in the slice.}
        \label{fig:Sparse_dist}
    \end{figure}

    \autoref{fig:Sparse_dist} emphasises the spatial proximity between the different maps by representing, for each $\boldsymbol{I}_J$, where $J$ denotes the fraction of galaxies we kept to compute the map, the cumulative distribution of $\{d_x^J\}_{x \in \boldsymbol{\Gamma}_{0.25}(\boldsymbol{I}_{100})}$ defined, for a position $x$ in the set $\boldsymbol{\Gamma}_{0.25}(\boldsymbol{I}_{100})$, as
    \begin{equation}  \label{eq:Proximity}
        d_x^J = \min_{x' \in \boldsymbol{\Gamma}_{0.25}(\boldsymbol{I}_{J})} \rVert x - x' \lVert_2.
    \end{equation}
    Hence $d_x^J$ corresponds to the closest distance from a position $x$ in the original skeleton obtained by keeping all subhalos, namely $\boldsymbol{\Gamma}_{0.25}(\boldsymbol{I}_{100})$, to a given thresholded map $\boldsymbol{\Gamma}_{0.25}(\boldsymbol{I}_J)$. This way, the distribution of $d_x^J$ measures how far the original pattern is from the one obtained with $J\%$ of the data points.
    
    In more than $95\%$ of the cases, the original pattern finds a closest point in the $83\%$ and the $60\%$ maps at less than $1.8$Mpc/h, showing that structures found in the three maps are spatially close and about the thickness of typical filaments \citep{Cautun2014}.
    When $M^{\text{cut}}$ increases, the filamentary pattern traces the most prominent parts of the structure with a loss of some small scales and hence highlights coarser and coarser structures. Even though the pattern is rough with only $31\%$ of the data points used, we still observe a nice correlation with previous maps highlighting coherent structures with $90\%$ of the original pattern being retrieved at less than $3$Mpc/h. As expected, an unrealistic scenario where we use only $10\%$ of the data points associated with the most massive subhalos degrades the reconstruction of the filamentary pattern. Yet, the recovered structures show a coarse but coherent connectivity between regions. This illustrates the ability of T-ReX to recover the underlying structure with high stability with respect to deformation of the input distribution of data points.
    
    \subsection{Application to 3D data} \label{Comparison}
    In this section, we apply T-ReX on the 3D distribution of halos obtained from a Gadget-2 simulation (see \sectRefShort{Data}) and compare our results with some other existing procedures that have also been run on the same dataset.
    Although the original review \citep{Libeskind2017} considers over a dozen different methods, we focus the comparison on three procedures, namely Nexus+, DisPerSE and Bisous, so that we have a broad set of different methods using respectively scale-space representation, topological considerations, or stochastic approach to recover the filamentary pattern. Nexus+ \citep{Nexus} is a classification algorithm inspired by image processing and based on filtering techniques leading to state-of-the-art environment classification able to identify clusters, filaments and walls. The main idea is to assume that the local morphology of the density field fully encodes the environmental information. Eigenvalues of the Hessian of the density field are thus used to compute an environmental signature in each voxel of the smoothed field. The key ingredient is to compute this signature for a set of smoothed fields with a log-Gaussian filter over a range of different scales to highlight structures of different sizes. Physically motivated criteria are then used to threshold signature values and attribute a classification to each volume element. Bisous \citep{Stoica2007} is a publicly available\footnote{\url{https://www.ascl.net/1512.008}} stochastic method based on halos positions aiming at identifying the filamentary structure using a set of random parametric cylinders. Filaments are modelled as aligned and contiguous small cylinders of a given size in the galaxy distribution.
    The Bisous model generates two maps allowing to extract filaments spine; one characterising the probability to find a filament at a given position called the visit map and an other one corresponding to the filament orientation field. This way, spines are defined as dense regions and are aligned with the axis of the different cylinders.
    
    We note that not only do these methods have very different mathematical definitions for what they all call clusters, filaments, and walls, but they also have been run with different input, using either DM particles or halos.
    
    We applied T-ReX to the full halo distribution of the 3D simulated box ($281465$ halos in total) and built a $100\times 100\times 100$ grid map like other methods. For T-ReX, this means that the final probability map is computed over a $100^3$ grid in which all visited voxels are considered part of the filamentary structure. As T-ReX is using 1D objects (segments of the RMST) sampled over the input space, it is preferable, for illustration and comparison, to give its filamentary pattern a 'thickness' by smoothing the obtained probability map. Whenever a voxel is classified as part of the filamentary structure, a smoothing is thus performed over its 26 direct neighbors. In what follows, we call this version T-ReX$_s$ while the original result is referred to as T-ReX$_{us}$.
    
\begin{figure*}
        \centering
        \includegraphics[width=0.95\linewidth]{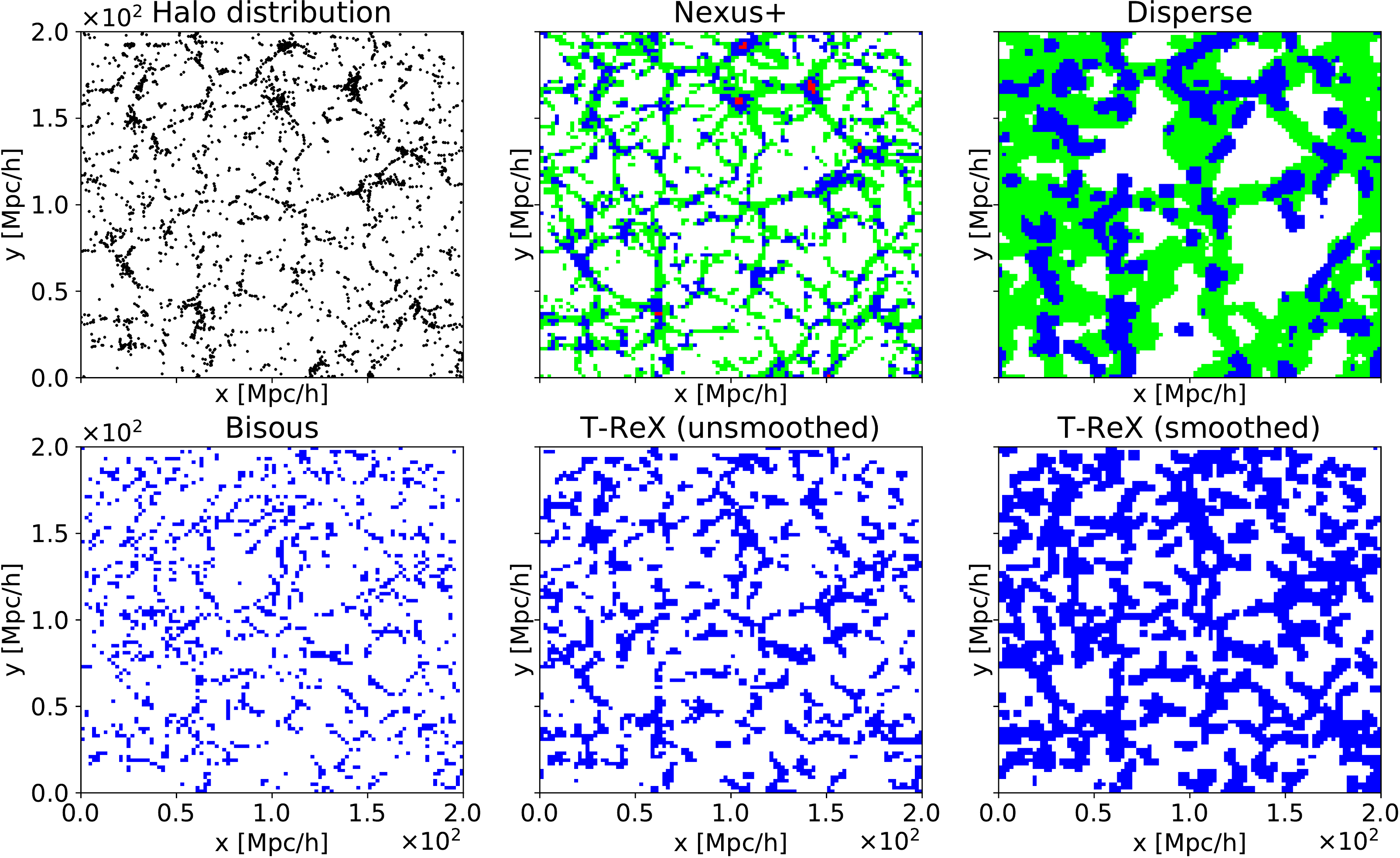}
        
        \medskip
        
        \caption{Identification results provided by four detection methods on a randomly chosen $2$Mpc/h depth slice of the full 3D detection for each method. Green pixels are walls, blue are filaments, red are clusters and white are voids or unclassified regions.}
        \label{fig:3D_comparison}
    \end{figure*}

    For illustration, following \cite{Libeskind2017}, we show in \figRefShort{fig:3D_comparison} the results of the classification provided by each method for a $2$Mpc/h depth slice from which FoF halos were extracted (top left panel of \figRefShort{fig:3D_comparison}). We note that all methods have been run over the full 3D cube and this is a projected slice of the detection. It is also worth noting that T-ReX identifies the filamentary pattern as a whole and does not classify the environment into clusters, filaments and walls as Nexus and DisPerSE do. To perform the comparison, we must look at the full pattern provided by each method and compare it with our extracted skeleton.
    We observe that T-ReX provides a satisfactory connectivity of the halos through the slice. In its smoothed version, it leads to thicker filaments compared to the results of Nexus+ and Bisous but thinner ones than Disperse, and retrieves most of the structures (filaments, walls, and clusters) obtained by the Nexus+ algorithm.
    
    Even though these methods have been developed with different approaches, it is interesting to see whether they agree or not in the detection of the filamentary pattern. To do so in a quantitative way, we could use the proximity measurement of Eq. \eqref{eq:Proximity} but as the resulting patterns are presented on a $2$Mpc/h grid, the distance between them would not be accurate. Hence, we introduce a similarity measurement as follows:\ considering the answers provided by two detection methods, $H_1$ and $H_2$, such that $H_{\bullet} (x) = 1$ if the position $x$ is part of the filamentary structure and $0$ otherwise, the similarity measurement is defined as:
    \begin{equation} \label{eq:Similarity}
        \mathcal{S}(H_1, H_2) = \frac{\abs{H_1 \cap H_2}}{\abs{H_1}},
    \end{equation}
    where $\abs{H_i}$ denotes the cardinal of $H_i$ defined as $\sum_{x} 1_{H_i(x) = 1}$ and $\abs{H_1 \cap H_2}$ is the cardinal of the intersection between $H_1$ and $H_2$ detections defined as $\sum_{x} 1_{H_1(x) = 1} 1_{H_2(x) = 1}$.
    Hence, $\mathcal{S}(H_1, H_2)$ measures the proportion of $H_1$ detections that are contained in $H_2$ and is thus asymmetric. In other words, if we consider $H_2$ as a reference, $\mathcal{S}(H_1, H_2)$ represents the proportion of true detections provided by $H_1$. Of course, such a simple metric does not provide the full information on the similarity between the considered patterns. This measure must then be completed in tandem with others, or with visual inspection, as we have done here.
    
    \autoref{tab:Similarity} shows the similarity indices between all considered methods for the entire 3D cube. We observe that $85\%$ of the detections provided by the unsmoothed version of T-ReX are contained in the Nexus+ skeleton and $81\%$ of the Nexus+ detections are found by the smoothed version of T-ReX. This indicates that the smoothed version of T-Rex contains a large part of the Nexus+ skeleton but with a larger amount of the volume detected, explained by the smoothing leading to a thicker filamentary pattern. The same tendency is observed concerning Bisous for which the detections are mostly contained in other skeletons (last row of \autoref{tab:Similarity}) but not reciprocally (last column of \autoref{tab:Similarity}). This is due to the sparse and unconnected detection provided by the Bisous method. The thick skeleton of DisPerSE also tends to contain a large fraction of other skeletons (fourth column of \autoref{tab:Similarity}) but it fills so much volume, which is not contained in the latter (fourth line of \autoref{tab:Similarity}).
    
    \begin{table}
        \centering
        \caption{Index of similarity $\mathcal{S}(H_1, H_2)$ as defined in eq. \eqref{eq:Similarity} between the considered methods applied on the entire 3D cube. T-ReX$_{us}$ refers to the unsmoothed version of the detection while T-ReX$_{s}$ refers to the smoothed one over the 26 neighboring voxels.}
        \label{tab:Similarity}
        
        \smallskip
        
        \begin{tabular}{l|*{5}r}
        \toprule
        \diagbox[width=2.80cm, height=2.80cm]{\raisebox{5pt}{\hspace*{0.25cm}$H_1$}}{\raisebox{-.67cm}{\rotatebox{90}{$H_2$}}} & \raisebox{-0.95cm}{\rotatebox{90}{T-ReX$_{us}$}} & \raisebox{-0.95cm}{\rotatebox{90}{T-ReX$_s$}} & \raisebox{-0.95cm}{\rotatebox{90}{Nexus+}} & \raisebox{-0.95cm}{\rotatebox{90}{DisPerSE}} & \raisebox{-0.95cm}{\rotatebox{90}{Bisous}} \\
        \midrule \\
        \hspace*{0.25cm}T-ReX$_{us}$ & 1 & 1 & 0.85 & 0.62 & 0.37 \\ \\
        \hspace*{0.25cm}T-ReX$_s$ & 0.48 & 1 & 0.62 & 0.62 & 0.24 \\ \\
        \hspace*{0.25cm}Nexus+ & 0.53 & 0.81 & 1 & 0.62 & 0.30 \\ \\
        \hspace*{0.25cm}DisPerSE & 0.22 & 0.46 & 0.35 & 1 & 0.12 \\ \\
        \hspace*{0.25cm}Bisous & 0.66 & 0.87 & 0.86 & 0.62 & 1 \\ \\
        \bottomrule
        \end{tabular}%
    \end{table}%
    
\section{Conclusion}

In this paper, we present T-ReX, a graph-based algorithm aimed at an automatic retrieval of the underlying density from a discrete set of points. We show that it can be used to uncover the natural filamentary pattern of the Cosmic Web from a 2D or 3D galaxy distribution.
The key idea of T-ReX is to find a set of centroids paving a given set of data points in its ridges by enforcing a predefined topology. To do so, the minimum spanning tree is computed over those centroids, which are iteratively moved to obtain a smoothed version of the MST. To characterise the reliability of the underlying filamentary structure, a without replacement bootstrap is used where several regularised MST are computed over a subset of data points chosen randomly and uniformly. In this way, we can build a probability map of those realisations to get the most frequent paths and highlight some regions as being part of the underlying filamentary pattern with high reliability.

For the sake of simplicity and because this topology is, at first, a fitting representation of the filamentary structure, we chose the tree topology for the centroids to highlight ridges of the point cloud distribution of galaxies. In addition, the MST provides a natural way to connect observed data points with the possibility to infer the underlying filamentary pattern by minimising the total distance linking them. However, the presented framework (see \sectRefShort{RGMM}) is more general and can use any kind of graph construction. Hence, it could be interesting to investigate other topologies and in other contexts that detecting ridges. In particular, its nearest neighbors have been recently applied in several cosmological studies, such as \cite{Coutinho2016}, to find new metrics characterising the Cosmic Web using graphs. Also, studying the properties of the regularised tree representation in the same way as it is done for the usual MST \citep[see e.g.][]{Colberg2007, Naidoo2019} could be of interest.

In this paper, we mainly focus the application of the procedure on simulated datasets. When dealing with real data, in addition to the mathematical considerations of defining and extracting the filamentary pattern, we face the usual technical issues of observed data: noise, outliers, uneven distribution of the samples, sparsity of the representation, selection, and observational effects. Even though we showed some robustness of the estimate to noise and outliers, the ability of minimum spanning tree methods to get rid of observational (redshift-space distortions) and selection effects (missing parts of the sky) in real cosmological surveys could be a consideration of further studies.


\section*{Acknowledgments}
The authors thank Pr. Einasto for helpful comments that improved the quality of the paper. We are grateful to the scikit-learn \citep{Pedregosa2011} team for making easier and sometimes straightforward the implementation and testing of parts of the algorithm. TB thanks Pr. S. White for fruitful discussions.
The authors also thank all members of the ByoPiC team\footnote{\url{https://byopic.eu/team}} for useful comments and discussions.
This research was supported by funding for the ByoPiC project from the European Research Council (ERC) under the European Union’s Horizon 2020 research and innovation program grant number ERC-2015-AdG 695561.

\medskip

\bibliographystyle{aa}
\bibliography{bibtex_200206}

\begin{thebibliography}{75}
\expandafter\ifx\csname natexlab\endcsname\relax\def\natexlab#1{#1}\fi

\bibitem[{Alpaslan {et~al.}(2014{\natexlab{a}})Alpaslan, Robotham, Driver,
  Norberg, Baldry, Bauer, Bland-hawthorn, Brown, Cluver, Colless, Foster,
  Hopkins, Van~kampen, Kelvin, Lara-Lopez, Liske, Lopez-Sanchez, Loveday,
  Mcnaught-Roberts, Merson, \& Pimbblet}]{Alpaslan2014a}
Alpaslan, M., Robotham, A.~S., Driver, S., {et~al.} 2014{\natexlab{a}}, Monthly
  Notices of the Royal Astronomical Society, 438, 177

\bibitem[{Alpaslan {et~al.}(2014{\natexlab{b}})Alpaslan, Robotham, Obreschkow,
  Penny, Driver, Norberg, Brough, Brown, Cluver, Holwerda, Hopkins, van Kampen,
  Kelvin, Lara-Lopez, Liske, Loveday, Mahajan, \& Pimbblet}]{Alpaslan2014}
Alpaslan, M., Robotham, A.~S., Obreschkow, D., {et~al.} 2014{\natexlab{b}},
  Monthly Notices of the Royal Astronomical Society: Letters, 440, 1

\bibitem[{Aragon-Calvo {et~al.}(2010)Aragon-Calvo, Weygaert, \&
  Jones}]{AragonCalvo2010}
Aragon-Calvo, M., Weygaert, R. V.~D., \& Jones, B. J.~T. 2010, Monthly Notices
  of the Royal Astronomical Society, 408, 2163

\bibitem[{Aragon-Calvo {et~al.}(2007)Aragon-Calvo, Jones, van~de Weygaert, M.,
  \& der Hulst}]{MMF}
Aragon-Calvo, M.~A., Jones, B. J.~T., van~de Weygaert, R., M., J., \& der
  Hulst, V. 2007, Astronomy and Astrophysics, 474, 315

\bibitem[{Arag{\'{o}}n-Calvo {et~al.}(2010)Arag{\'{o}}n-Calvo, Platen, {Van De
  Weygaert}, \& Szalay}]{Spineweb}
Arag{\'{o}}n-Calvo, M.~A., Platen, E., {Van De Weygaert}, R., \& Szalay, A.~S.
  2010, Astrophysical Journal, 723, 364

\bibitem[{Barrow {et~al.}(1985)Barrow, Bhavsar, \& Sonoda}]{Barrow1985}
Barrow, J.~D., Bhavsar, S.~P., \& Sonoda, D.~H. 1985, Monthly Notices of the
  Royal Astronomical Society, 216, 17

\bibitem[{Bezdek(1981)}]{Bezdek1981}
Bezdek, J.~C. 1981, {Pattern Recognition with Fuzzy Objective Function
  Algorithms}, plenum pre edn. (Kluwer Academic Publishers), 267

\bibitem[{Bishop \& Svens{\'{e}}n(1998)}]{Bishop1998}
Bishop, C.~M. \& Svens{\'{e}}n, M. 1998, Neural Computation 10, 1, 215

\bibitem[{Bond {et~al.}(1996)Bond, Kofman, \& Pogosyan}]{Bond1996}
Bond, J.~R., Kofman, L., \& Pogosyan, D. 1996, Nature, 380, 603

\bibitem[{Bonjean {et~al.}(2018)Bonjean, Aghanim, Salom{\'{e}}, Douspis, \&
  Beelen}]{Bonjean2017}
Bonjean, V., Aghanim, N., Salom{\'{e}}, P., Douspis, M., \& Beelen, A. 2018,
  Astronomy and Astrophysics, 609, A49

\bibitem[{Borůvka(1926)}]{Boruvka1926}
Borůvka, O. 1926, Pr{\'{a}}ce Moravsk{\'{e}} př{\'{i}}rodov{\v{e}}deck{\'{e}}
  spole{\v{c}}nosti, 3, 37

\bibitem[{Bos {et~al.}(2014)Bos, {Van De Weygaert}, Kitaura, \&
  Cautun}]{Bos2014}
Bos, E.~G., {Van De Weygaert}, R., Kitaura, F., \& Cautun, M. 2014, Proceedings
  of the International Astronomical Union, 11, 271

\bibitem[{Cautun {et~al.}(2013)Cautun, van~de Weygaert, \& Jones}]{Nexus}
Cautun, M., van~de Weygaert, R., \& Jones, B.~J. 2013, Monthly Notices of the
  Royal Astronomical Society, 429, 1286

\bibitem[{Cautun {et~al.}(2014)Cautun, Weygaert, Jones, \& Frenk}]{Cautun2014}
Cautun, M., Weygaert, R. V.~D., Jones, B. J.~T., \& Frenk, C.~S. 2014, Monthly
  Notices of the Royal Astronomical Society, 441, 2923

\bibitem[{Chen {et~al.}(2015)Chen, Ho, Freeman, Genovese, \&
  Wasserman}]{Chen2015}
Chen, Y.~C., Ho, S., Freeman, P.~E., Genovese, C.~R., \& Wasserman, L. 2015,
  Monthly Notices of the Royal Astronomical Society, 454, 1140

\bibitem[{Codis {et~al.}(2018)Codis, Pogosyan, \& Pichon}]{Codis2018}
Codis, S., Pogosyan, D., \& Pichon, C. 2018, Monthly Notices of the Royal
  Astronomical Society, 479, 973

\bibitem[{Colberg(2007)}]{Colberg2007}
Colberg, J.~M. 2007, Monthly Notices of the Royal Astronomical Society, 375,
  337

\bibitem[{Coutinho {et~al.}(2016)Coutinho, Hong, Albrecht, Dey, Barab{\'{a}}si,
  Torrey, Vogelsberger, \& Hernquist}]{Coutinho2016}
Coutinho, B.~C., Hong, S., Albrecht, K., {et~al.} 2016, arXiv e-prints
  [\eprint[arXiv]{1604.03236}]

\bibitem[{de~Graaf {et~al.}(2019)de~Graaf, Cai, Heymans, \&
  Peacock}]{Degraaf2019}
de~Graaf, A., Cai, Y.-c., Heymans, C., \& Peacock, J.~A. 2019, Astronomy and
  Astrophysics, 624, A48

\bibitem[{Dempster {et~al.}(1977)Dempster, Laird, \& Rubin}]{Dempster1977}
Dempster, A.~P., Laird, N.~M., \& Rubin, D.~B. 1977, Journal of the Royal
  Statistical Society, 39, 1

\bibitem[{Dietrich {et~al.}(2012)Dietrich, Werner, Clowe, Finoguenov, Kitching,
  Miller, \& Simionescu}]{Dietrich2012}
Dietrich, J.~P., Werner, N., Clowe, D., {et~al.} 2012, Nature, 487, 202

\bibitem[{Doroshkevich \& Shandarin(1978)}]{Shandarin1978}
Doroshkevich \& Shandarin. 1978, Sovast, 22, 653

\bibitem[{Dubois {et~al.}(2014)Dubois, Pichon, Welker, Borgne, Devriendt,
  Laigle, Codis, Pogosyan, Arnouts, Benabed, Bertin, Blaizot, Bouchet, Cardoso,
  Colombi, Lapparent, Desjacques, Gavazzi, Kassin, Kimm, Mccracken, Milliard,
  Peirani, Prunet, Rouberol, Silk, Slyz, Sousbie, Teyssier, Tresse, Treyer,
  Vibert, \& Volonteri}]{Dubois2014}
Dubois, Y., Pichon, C., Welker, C., {et~al.} 2014, Monthly Notices of the Royal
  Astronomical Society, 444, 1453

\bibitem[{Durbin \& Willshaw(1987)}]{Durbin1987}
Durbin, R. \& Willshaw, D. 1987, Nature, 326, 14

\bibitem[{Eckert {et~al.}(2015)Eckert, Jauzac, Shan, Kneib, Erben, Israel,
  Jullo, Klein, Massey, Richard, \& Tchernin}]{Eckert2015}
Eckert, D., Jauzac, M., Shan, H., {et~al.} 2015, Nature, 528, 105

\bibitem[{Edelsbrunner {et~al.}(2002)Edelsbrunner, Letscher, \&
  Zomorodian}]{Edelsbrunner2002}
Edelsbrunner, H., Letscher, D., \& Zomorodian, A. 2002, Discrete Computational
  Geometry, 28, 511

\bibitem[{Einasto {et~al.}(1980)Einasto, Joeveer, \& Saar}]{Einasto1980}
Einasto, J., Joeveer, M., \& Saar, E. 1980, Monthly Notices of the Royal
  Astronomical Society, 193, 353

\bibitem[{Epps \& Hudson(2017)}]{Epps2017}
Epps, D. \& Hudson, M.~J. 2017, Monthly Notices of the Royal Astronomical
  Society, 468, 2605

\bibitem[{Forman(1998)}]{Forman1998}
Forman, R. 1998, Advances in Mathematics, 145, 90

\bibitem[{Genovese {et~al.}(2014)Genovese, Perone-Pacifico, Verdinelli, \&
  Wasserman}]{Genovese2014}
Genovese, C.~R., Perone-Pacifico, M., Verdinelli, I., \& Wasserman, L. 2014,
  Annals of Statistics, 42, 1511

\bibitem[{Gheller \& Vazza(2019)}]{Gheller2019}
Gheller, C. \& Vazza, F. 2019, Monthly Notices of the Royal Astronomical
  Society, 486, 981

\bibitem[{Gheller {et~al.}(2016)Gheller, Vazza, Br, Alpaslan, Holwerda,
  Hopkins, \& Liske}]{Gheller2016}
Gheller, C., Vazza, F., Br, M., {et~al.} 2016, Monthly Notices of the Royal
  Astronomical Society, 462, 448

\bibitem[{Gorban \& Zinovyev(2005)}]{Elmap2005}
Gorban, A. \& Zinovyev, A. 2005, Computing, 75, 359

\bibitem[{Gouin {et~al.}(2017)Gouin, Gavazzi, Codis, Pichon, Peirani, \&
  Dubois}]{Gouin2017}
Gouin, C., Gavazzi, R., Codis, S., {et~al.} 2017, Astronomy and Astrophysics,
  605, A27

\bibitem[{Hastie {et~al.}(1989)Hastie, Stuetzle, Hastie, \&
  Stuetzle}]{Hastie1989}
Hastie, T., Stuetzle, W., Hastie, T., \& Stuetzle, W. 1989, Journal of the
  American Statistical Association, 84, 502

\bibitem[{H{\'{e}}bert-Dufresne {et~al.}(2016)H{\'{e}}bert-Dufresne, Grochow,
  \& Allard}]{Hebert-Dufresne2016}
H{\'{e}}bert-Dufresne, L., Grochow, J.~A., \& Allard, A. 2016, Scientific
  Reports, 6, 1

\bibitem[{Jasche \& Wandelt(2013)}]{Jasche2013}
Jasche, J. \& Wandelt, B.~D. 2013, Monthly Notices of the Royal Astronomical
  Society, 432, 894

\bibitem[{Joeveer {et~al.}(1978)Joeveer, Einasto, \& Tago}]{Joeveer1978}
Joeveer, M., Einasto, J., \& Tago, E. 1978, Monthly Notices of the Royal
  Astronomical Society, 185, 357

\bibitem[{Kitaura(2013)}]{Kitaura2012}
Kitaura, F.-s. 2013, Monthly Notices of the Royal Astronomical Society, 429,
  L84

\bibitem[{Kraljic {et~al.}(2020)Kraljic, Dav{\'{e}}, \& Pichon}]{Kraljic2019}
Kraljic, K., Dav{\'{e}}, R., \& Pichon, C. 2020, Monthly Notices of the Royal
  Astronomical Society, 237

\bibitem[{Kullback \& Leibler(1951)}]{KL1951}
Kullback, S. \& Leibler, R. 1951, Annals of Mathematical Statistics, 22, 79

\bibitem[{Kuutma {et~al.}(2017)Kuutma, Tamm, \& Tempel}]{Kuutma2017}
Kuutma, T., Tamm, A., \& Tempel, E. 2017, Astronomy and Astrophysics, 600, L6

\bibitem[{Laigle {et~al.}(2018)Laigle, Pichon, Arnouts, McCracken, Dubois,
  Devriendt, Slyz, {Le Borgne}, Benoit-L{\'{e}}vy, Hwang, Ilbert, Kraljic,
  Malavasi, Park, \& Vibert}]{Laigle2018}
Laigle, C., Pichon, C., Arnouts, S., {et~al.} 2018, Monthly Notices of the
  Royal Astronomical Society, 474, 5437

\bibitem[{Leclercq {et~al.}(2016)Leclercq, Lavaux, Jasche, \&
  Wandelt}]{Leclercq2016}
Leclercq, F., Lavaux, G., Jasche, J., \& Wandelt, B. 2016, Journal of Cosmology
  and Astroparticle Physics, 2016, 1

\bibitem[{Libeskind {et~al.}(2017)Libeskind, van~de Weygaert, Cautun, Falck,
  Tempel, Abel, Alpaslan, Arag{\'{o}}n-Calvo, Forero-Romero, Gonzalez,
  Gottl{\"{o}}ber, Hahn, Hellwing, Hoffman, Jones, Kitaura, Knebe, Manti,
  Neyrinck, Nuza, Padilla, Platen, Ramachandra, Robotham, Saar, Shandarin,
  Steinmetz, Stoica, Sousbie, \& Yepes}]{Libeskind2017}
Libeskind, N.~I., van~de Weygaert, R., Cautun, M., {et~al.} 2017, Monthly
  Notices of the Royal Astronomical Society, 473, 1195

\bibitem[{Lurie(1999)}]{Lurie1999}
Lurie, J. 1999, ACM SIGACT News, 30, 14

\bibitem[{Macqueen(1967)}]{Macqueen1967}
Macqueen, J. 1967, Mathematical reviews, 281

\bibitem[{Malavasi {et~al.}(2019)Malavasi, Aghanim, Tanimura, Bonjean, \&
  Douspis}]{Malavasi2019}
Malavasi, N., Aghanim, N., Tanimura, H., Bonjean, V., \& Douspis, M. 2019,
  arXiv e-prints [\eprint[arXiv]{1910.11879v2}]

\bibitem[{Malavasi {et~al.}(2017)Malavasi, Arnouts, Vibert, de~la Torre,
  Moutard, Pichon, Davidzon, Kraljic, Bolzonella, Guzzo, Garilli, Scodeggio,
  Granett, Abbas, Adami, Bottini, Cappi, Cucciati, Franzetti, Fritz, Iovino,
  Krywult, {Le Brun}, {Le F{\`{e}}vre}, Maccagni, Malek, Marulli, Polletta,
  Pollo, Tasca, Tojeiro, Vergani, Zanichelli, Bel, Branchini, Coupon, {De
  Lucia}, Dubois, Hawken, Ilbert, Laigle, Moscardini, Sousbie, Treyer, \&
  Zamorani}]{Malavasi2017}
Malavasi, N., Arnouts, S., Vibert, D., {et~al.} 2017, Monthly Notices of the
  Royal Astronomical Society, 465, 3817

\bibitem[{Mao {et~al.}(2016)Mao, Li, {Ivor W.}, \& Sun}]{Mao2016}
Mao, Q., Li, W., {Ivor W.}, T., \& Sun, Y. 2016, arXiv e-prints
  [\eprint[arXiv]{1512.02752v2}]

\bibitem[{Mao {et~al.}(2015)Mao, Yang, Wang, Goodison, \& Sun}]{Mao2015}
Mao, Q., Yang, L., Wang, L., Goodison, S., \& Sun, Y. 2015, Proceedings of the
  2015 SIAM International Conference on Data Mining, 792

\bibitem[{Martinez {et~al.}(2016)Martinez, Muriel, \& Coenda}]{Martinez2016}
Martinez, H., Muriel, H., \& Coenda, V. 2016, Monthly Notices of the Royal
  Astronomical Society, 445, 127

\bibitem[{Moccia {et~al.}(2018)Moccia, Momi, Hadji, \& Mattos}]{Moccia2018}
Moccia, S., Momi, E.~D., Hadji, S.~E., \& Mattos, L.~S. 2018, Computer methods
  and programs in biomedicine, 158, 71

\bibitem[{More {et~al.}(2011)More, Kravtsov, Dalal, \&
  Gottl{\"{o}}ber}]{More2011}
More, S., Kravtsov, A.~V., Dalal, N., \& Gottl{\"{o}}ber, S. 2011,
  Astrophysical Journal, Supplement Series, 195 [\eprint[arXiv]{1103.0005}]

\bibitem[{Naidoo {et~al.}(2019)Naidoo, Whiteway, Massara, Gualdi, Lahav, \&
  Viel}]{Naidoo2019}
Naidoo, K., Whiteway, L., Massara, E., {et~al.} 2019, Monthly Notices of the
  Royal Astronomical Society [\eprint[arXiv]{1907.00989v1}]

\bibitem[{Nicastro {et~al.}(2018)Nicastro, Kaastra, Krongold, Borgani,
  Branchini, Cen, Dadina, Danforth, Elvis, Fiore, Gupta, Mathur, Mayya,
  Paerels, Piro, Rosa-Gonzalez, Schaye, Shull, Torres-Zafra, Wijers, \&
  Zappacosta}]{Nicastro2018}
Nicastro, F., Kaastra, J., Krongold, Y., {et~al.} 2018, Nature, 558, 406

\bibitem[{Pedregosa {et~al.}(2011)Pedregosa, Weiss, \& Brucher}]{Pedregosa2011}
Pedregosa, F., Weiss, R., \& Brucher, M. 2011, Journal of Machine Learning
  Research, 12, 2825

\bibitem[{Qiu {et~al.}(2017)Qiu, Qi, Ying, Li, Raghav, Pliner, \&
  Trapnell}]{Qiu2017}
Qiu, X., Qi, M., Ying, T., {et~al.} 2017, Nature Methods, 14, 979

\bibitem[{Roweis \& Saul(2000)}]{Roweis2000}
Roweis, S.~T. \& Saul, L.~K. 2000, Science, 290, 2323

\bibitem[{Sarron {et~al.}(2019)Sarron, Adami, Durret, \& Laigle}]{Sarron2019}
Sarron, F., Adami, C., Durret, F., \& Laigle, C. 2019, Astronomy and
  Astrophysics, A49

\bibitem[{Schaap \& Weygaert(2000)}]{Schaap2000}
Schaap, W. \& Weygaert, R. 2000, Astronomy and Astrophysics, 363, L29

\bibitem[{Silverman(1986)}]{Silverman1986}
Silverman, B. 1986, Monographs on Statistics and Applied Probability

\bibitem[{Smola {et~al.}(2001)Smola, Mika, Sch, \& Williamson}]{Smola2001}
Smola, A.~J., Mika, S., Sch, B., \& Williamson, R.~C. 2001, The Journal of
  Machine Learning Research, 1, 179

\bibitem[{Sousbie(2011)}]{DisperseTheory}
Sousbie, T. 2011, Monthly Notices of the Royal Astronomical Society, 414, 350

\bibitem[{Springel {et~al.}(2008)Springel, Wang, Vogelsberger, Ludlow, Jenkins,
  Helmi, Navarro, Frenk, \& White}]{Springel2008}
Springel, V., Wang, J., Vogelsberger, M., {et~al.} 2008, Monthly Notices of the
  Royal Astronomical Society, 391, 1685

\bibitem[{Springel {et~al.}(2005)Springel, White, Jenkins, Frenk, Yoshida, Gao,
  Navarro, Thacker, Croton, Helly, Peacock, Cole, Thomas, Couchman, Evrard,
  Colberg, \& Pearce}]{Springel2005}
Springel, V., White, S. D.~M., Jenkins, A., {et~al.} 2005, Nature, 435, 629

\bibitem[{Stoica {et~al.}(2007)Stoica, Mart{\'{i}}nez, \& Saar}]{Stoica2007}
Stoica, R.~S., Mart{\'{i}}nez, V.~J., \& Saar, E. 2007, Journal of the Royal
  Statistical Society. Series C: Applied Statistics, 56, 459

\bibitem[{Tanimura {et~al.}(2019{\natexlab{a}})Tanimura, Aghanim, Bonjean,
  Malavasi, \& Douspis}]{Tanimura2019}
Tanimura, H., Aghanim, N., Bonjean, V., Malavasi, N., \& Douspis, M.
  2019{\natexlab{a}}, arXiv e-prints [\eprint[arXiv]{1911.09706v1}]

\bibitem[{Tanimura {et~al.}(2019{\natexlab{b}})Tanimura, Hinshaw, McCarthy,
  {Van Waerbeke}, Ma, Mead, Hojjati, \& Tr{\"{o}}ster}]{Tanimura2017}
Tanimura, H., Hinshaw, G., McCarthy, I.~G., {et~al.} 2019{\natexlab{b}},
  Monthly Notices of the Royal Astronomical Society, 483, 223

\bibitem[{Tibshirani(1992)}]{Tibshirani1992}
Tibshirani, R. 1992, Statistics and Computing, 2, 183

\bibitem[{Tibshirani {et~al.}(2001)Tibshirani, Walther, \&
  Hastie}]{Tibshirani2001}
Tibshirani, R., Walther, G., \& Hastie, T. 2001, Journal of the Royal
  Statistical Society. Series B: Statistical Methodology, 63, 411

\bibitem[{Vogelsberger {et~al.}(2014)Vogelsberger, Genel, Springel, Torrey,
  Sijacki, Xu, Snyder, Nelson, \& Hernquist}]{Vogelsberger2014}
Vogelsberger, M., Genel, S., Springel, V., {et~al.} 2014, Monthly Notices of
  the Royal Astronomical Society, 444, 1518

\bibitem[{York {et~al.}(2000)York, Adelman, Anderson, Anderson, Annis, \&
  Bahcall}]{York2000}
York, D.~G., Adelman, J., Anderson, J., {et~al.} 2000, Astrophysical Journal,
  120, 1579

\bibitem[{Yuille(1990)}]{Yuille1990}
Yuille, A.~L. 1990, Neural Computation, 2, 1

\bibitem[{Zel'dovich(1970)}]{Zeldovich1970}
Zel'dovich. 1970, Astronomy and Astrophysics, 500, 13

\end{thebibliography}

\end{document}